\documentclass[twocolumn]{aa}  
\usepackage{natbib,twoopt}
\usepackage{hyperref}
\hypersetup{colorlinks=true,linkcolor=blue,citecolor=blue,filecolor=magenta,urlcolor=cyan}
\bibpunct{(}{)}{;}{a}{}{,}             
\usepackage{graphicx}
\usepackage{txfonts}

\def\aap{{\em A}\&{\em A}}

\def\apjl{{\em ApJL}}
\def\apjs{{\em ApJS}}

\newcommand{\kms}{\ifmmode {\rm km\ s}^{-1} \else km s$^{-1}$\ \fi}
\newcommand{\ergs}{\ifmmode {\rm erg\ s}^{-1} \else erg s$^{-1}$\ \fi}
\newcommand{\lb}{\ifmmode L_{\rm Bol} \else $L_{\rm Bol}$\ \fi}
\newcommand{\ledd}{\ifmmode L_{\rm Edd} \else $L_{\rm Edd}$\ \fi}
\newcommand{\lx}{\ifmmode L_{\rm 2-10keV} \else  $L_{\rm 2-10keV}$\ \fi}
\newcommand{\ha}{\hbox{H$\alpha$}}
\newcommand{\hb}{\hbox{H$\beta$}}

\newcommand{\mbh}{\ifmmode M_{\rm BH}  \else $M_{\rm BH}$\ \fi}
\newcommand{\lv}{\ifmmode \lambda L_{\lambda}(5100\Ang) \else $\lambda L_{\lambda}(5100\Ang)$\ \fi}
\newcommand{\lbol}{\ifmmode L_{\rm Bol} \else $L_{\rm Bol}$\ \fi}

\newcommand{\oi}{\hbox{[O\,{\sc i}]}}
\newcommand{\oii}{\hbox{[O\,{\sc ii}]}}
\newcommand{\nii}{\hbox{[N\,{\sc ii}]}}
\newcommand{\sii}{\hbox{[S\,{\sc ii}]}}
\newcommand{\oiii}{\hbox{[O\,{\sc iii}]}}

\newcommand{\msun}{ M_{\odot}}

\newcommand{\hii}{\ion{H}{II}}

\newcommand{\oh}{\ifmmode 12+ \log({\rm O/H}) \else 12+log(O/H) \fi}
\newcommand{\mdot}{\ifmmode \dot{m} \else \dot{m} \fi }
\newcommand{\llog}{\ifmmode {\rm log} \else {\rm log} \fi }
\newcommand{\Ang}{\mathring{\mathrm{A}}}

\begin{document}

   \titlerunning{Merger-induced star formation in NGC 4809/4810}
   \authorrunning{Gao et al. 2023}

   \title{Merger-induced star formation in low-metallicity dwarf galaxy NGC 4809/4810}

   \author{Yulong Gao\inst{1,2}, Qiusheng Gu\inst{1,2}, Guilin Liu\inst{3,4}, Hongxin Zhang\inst{3,4}, Yong Shi\inst{1,2}, Jing Dou\inst{1,2}, Xiangdong Li\inst{1,2} and Xu Kong\inst{3,4,5}
          }

   \institute{School of Astronomy and Space Science, Nanjing University, Nanjing 210093, China \ \email{yulong@nju.edu.cn}
             \and
             Key Laboratory of Modern Astronomy and Astrophysics (Nanjing University), Ministry of Education, Nanjing 210093, China
             \and
             CAS Key Laboratory for Research in Galaxies and Cosmology, Department of Astronomy, University of Science and Technology of China, Hefei 230026, China
             \and 
             School of Astronomy and Space Science, University of Science and Technology of China, Hefei 230026, China
             \and 
             Frontiers Science Center for Planetary Exploration and Emerging Technologies, University of Science and Technology of China, Hefei, Anhui, 230026, China
             }


 
  \abstract
   {The physical mechanisms driving starbursts in dwarf galaxies are unclear, and the effects of mergers on star formation in these galaxies are still uncertain.}
   {We explore how the merger process affects star formation in metal-poor dwarf galaxies by analyzing high-spatial-resolution ($\sim$ 70 pc) integral field spectrograph observations of ionized gas.}
   {We use archival data from the Very Large Telescope/Multi Unit Spectroscopic Explorer to map the spatial distribution of strong emission lines (e.g., $\hb$, $\ha$, $\oiii\lambda5007$, $\nii\lambda6583$, etc) in the nearby merging star-forming dwarf galaxy system NGC 4809/4810.}
   {We identify approximately 112 star-forming knots scattered among the two galaxies, where the gas-phase metallicity distribution is inhomogeneous and mixing with metal-poor and metal-rich ionized gas. Star-forming knots at the interacting region show lower metallicity, the highest star formation rates (SFRs) and SFR to resolved main-sequence-relation (rMSR) ratios. Ionized gas exhibits an obvious northeast-southwest velocity gradient in NGC 4809, while seemingly mixed in NGC 4810. High virial parameters and the stellar mass-size relation of $\hii$ regions indicate that these regions are dominated by direct radiation pressure from massive stars/clusters and persistently expanding. We find two different stellar mass surface density-stellar age relations in NGC 4809 and NGC 4810, and the stellar ages of NGC 4810 are systematically younger than in NGC 4809.}
   {Our study suggests that the merging stage of two dwarf galaxies can induce starburst activities at the interaction areas, despite the metal-deficient environment. Considering the high specific SFRs and different stellar ages, we propose that the interaction initially triggered star formation in NGC 4809 and then drove star formation in NGC 4810.} 

   \keywords{Galaxies:ISM -- Galaxies: star formation -- Galaxies: individual: NGC 4809, NGC 4810 -- Galaxies: interactions  }
   \maketitle
%
\section{Introduction}

The physical mechanisms to trigger galaxy starbursts are not yet fully understood. In the star formation rate (SFR) vs. stellar mass ($M_*$) diagram, galaxies fall into two broad categories: star-forming galaxies (SFGs, or blue clouds), which are mostly spiral or late-type galaxies that follow an almost linear relationship between the SFR and the stellar mass \citep[known as the main sequence, MSR, ][]{Brinchmann2004a, Daddi2007, Peng2010, Lilly2013, Speagle2014}; and starburst galaxies, which experience an exceptionally intense phase of star formation and lie above the MSR \citep{elbaz2018, Orlitova2020}. Quiescent galaxies (QGs, or "red sequences"), mostly consisting of elliptical or early-type galaxies, have much lower SFR than SFGs and lie below the main sequence. Some "green valley" galaxies lie between the blue SFGs and red QGs and have been classified as transition zone or quenching galaxies \citep{Bell2003, Fang2012}. The existence of galaxies above the main sequence suggests that SFGs may boost their star formation at some point in their lifespan.

Galaxy mergers are one of the inevitable processes in galaxy evolution \citep{Hopkins2009}. Mergers disrupt gas rotation, leading to rapid gas inflow that fuels intense starbursts or feeds central massive black holes. Energetic feedback from active galactic nuclei (AGNs) or starbursts heats the interstellar medium (ISM), preventing gas from cooling and expelling gas from the host galaxy \citep{fabian2012, Cheung2016May,harrison2018}, thereby suppressing star formation and transforming blue SFGs into quiescent red massive elliptical galaxies. In the local Universe, major mergers mainly occur between spiral galaxies, which can result in ultraluminous infrared galaxies \citep[(U)LIRGs; ][]{Papadopoulos2007, Israel2015, Espada2018, Spence2018, Shangguan2019a}.

{How star formation is triggered} in dwarf galaxies with $M_* < 5 \times 10^9 \ \msun$ \citep{Stierwalt2015} remains a mystery. While more than 70\% of the galaxies in the local Universe are dwarfs, only a small percentage of them are starburst galaxies. {\cite{elbaz2018} identified the starburst galaxies as two different regimes. One is the global starbursts located above the MSR, showing high gas fraction and short gas depletion time. Another is these galaxies located within the scatter of MSR while containing some compact star formation regions with short gas depletion time ($\sim$ 150 Myr). Global starbursts preferentially occur in dwarfs, while the starbursts are most often located in the circumnuclear region in the massive galaxies, e.g., ultra luminous infrared galaxies ((U)LIRGs).} The dwarf starburst galaxies include blue compact galaxies (BCGs), Lyman-$\alpha$ reference sample \citep[LARS, ][]{hayes2013,ostlin2014}, Lyman-break analogs \citep[LBAs, ][]{heckman2001,heckman2005}, and Green Peas \citep[see the review in][]{Orlitova2020}. Although many BCGs are relatively isolated, their morphologies and velocity distribution suggest they've recently interacted with neighbors. Moreover, the stellar components in BCGs contain both young and old populations \citep{kunth2000}, indicating that interactions may trigger the starburst activity. However, previous star formation models suggest that due to the weak gravitational potential and strong turbulence of stellar winds, the conversion from molecular gas to stars may be inefficient. Thus, it remains an open question whether the merging of two dwarf galaxies can trigger starburst activity. Recent studies by \cite{Zhang2020h} and \cite{Zhang2020g} report an enhanced star cluster formation rate ($\propto$ SFR) in the galaxy VCC 848, a remnant of a gas-rich dwarf-dwarf merger, by approximately 1.0 dex during the past 1 Gyr relative to its earlier times. Meanwhile, using high-spatial-resolution observations from the Very Large Telescope (VLT)/Multi Unit Spectroscopic Explorer (MUSE) and Atacama Large Millimeter Array (ALMA), \cite{gao2022a} found that the post-merger galaxy Haro 11 is undergoing efficient molecular gas consumption and stellar mass assembly. However, previous studies rarely examine the detailed star formation activities of dwarf galaxies at pre-merger and merging stages.

This paper focuses on studying the ionized gas properties of NGC 4809/4810, a merging system in the nearby universe \citep[e.g.,][]{casasola2004a,Paudel2018}. NGC 4809 and NGC 4810 are currently colliding and show a small overlapping region. We obtained the redshift ($z$) of 0.00326 from the NASA/IPAC Extragalactic Database (NED) \footnote{\url{http://ned.ipac.caltech.edu}}, which corresponds to a luminosity distance of 14.0 Mpc and a scale of 67 pc per arc second. \cite{zou2019} used photometric data from the DESI Legacy Imaging survey ($g, r, z$ bands) \citep{dey2019} and the unWISE survey ($W1, W2$ bands) \citep{mainzer2014} to perform stellar population synthetic fitting and determine the total stellar mass of NGC 4809/4810 is around $2.5 \times 10^8 \ \msun$, indicating that it is a merging dwarf galaxy system. Single spectrum observations prevented a detailed analysis of the physical properties (density, temperature, dust extinction, and metallicity) of the ionized gas, the kinematics of stellar and ionized gas, and the effect of the merger on star formation activities. To address these questions, we collected and analyzed high-spatial-resolution integral field spectrograph (IFS) observation data by the VLT/MUSE to investigate the ionized gas properties and kinematics in NGC 4809/4810 at 67 pc scales.

This paper is organized as follows: In Section 2, we present the observations and data reduction and then derive the ionized gas properties. The main results and discussion are presented in Sections 3 and 4, respectively. We summarize our findings in Section 5. Throughout the paper, we assume a flat $\Lambda$CDM cosmology model with $\Omega_\Lambda=0.7$, $\Omega_{\rm m}=0.3$, and $H_0=70$ km s$^{-1}$ Mpc$^{-1}$. We adopt the solar metallicity ($Z_{\odot}$) as $\oh = 8.69$ \citep{Allende2001}.

\section{Observations and data reduction}
\label{sec:data}

%
\subsection{VLT/MUSE data}
\label{subsec:muse_data}

NGC 4809/4810 was observed by VLT/MUSE in May 2015 (ID: 095.D-0172; PI: Kuncarayakti) with an integration exposure time of approximately 0.8 hours on the two-component field-of-view regions. The fully reduced data cube was obtained from the ESO archive website\footnote{\url{http://archive.eso.org/scienceportal/home}}. The observations had a seeing value of about 0.5 $\arcsec$, and the full width at half maximum (FWHM) of the final image was approximately 0.8$\arcsec$. The rest-frame spectral range was $\rm 4750 - 9160 \Ang$ with a channel width of 1.25$\rm \Ang$.

To derive the flux of pure emission lines, we use the STARLIGHT package \citep{CidFernandes2005} to reproduce the stellar continuum. In this process, we assume the \cite{Chabrier2003} initial mass function (IMF) and perform a combination of 45 single stellar populations (SSPs) from the \cite{Bruzual2003} model, which consists of three different metallicities and 15 stellar ages. The SSP fitting results allowed us to derive the stellar mass and stellar age within each spatial pixel (spaxel) with an uncertainty of 0.11 dex and 0.14 dex \citep{Bruzual2003, CidFernandes2005}, respectively. We estimate the total stellar mass ($M_* \sim 2.4 \times 10^8 \ \msun$) by integrating the values in each pixel with a continuum S/N larger than 3.0, in which the stellar masses of NGC 4809 and NGC 4810 are about $1.3 \times 10^8 \ \msun$ and $1.1 \times 10^8 \ \msun$, respectively. After subtracting the stellar continuum synthesis, we apply multiple Gaussians to fit strong emission lines such as $\ha$, $\hb$, $\oiii\lambda\lambda4959,5007$, $\nii\lambda\lambda6548,6583$, and $\sii\lambda\lambda6717,6731$. To ensure reliable measurement of SFR and metallicity, we only consider spaxels with S/N($\ha,\hb,\oiii\lambda\lambda4959,5007$) $>$ 5 and S/N($\nii\lambda6583,\sii\lambda\lambda6717,6731$) $>$ 3. 

\subsection{Star formation knots}
\label{subsec:knots}

\begin{figure*}[t]
   \centering
   \includegraphics[width=0.45\textwidth]{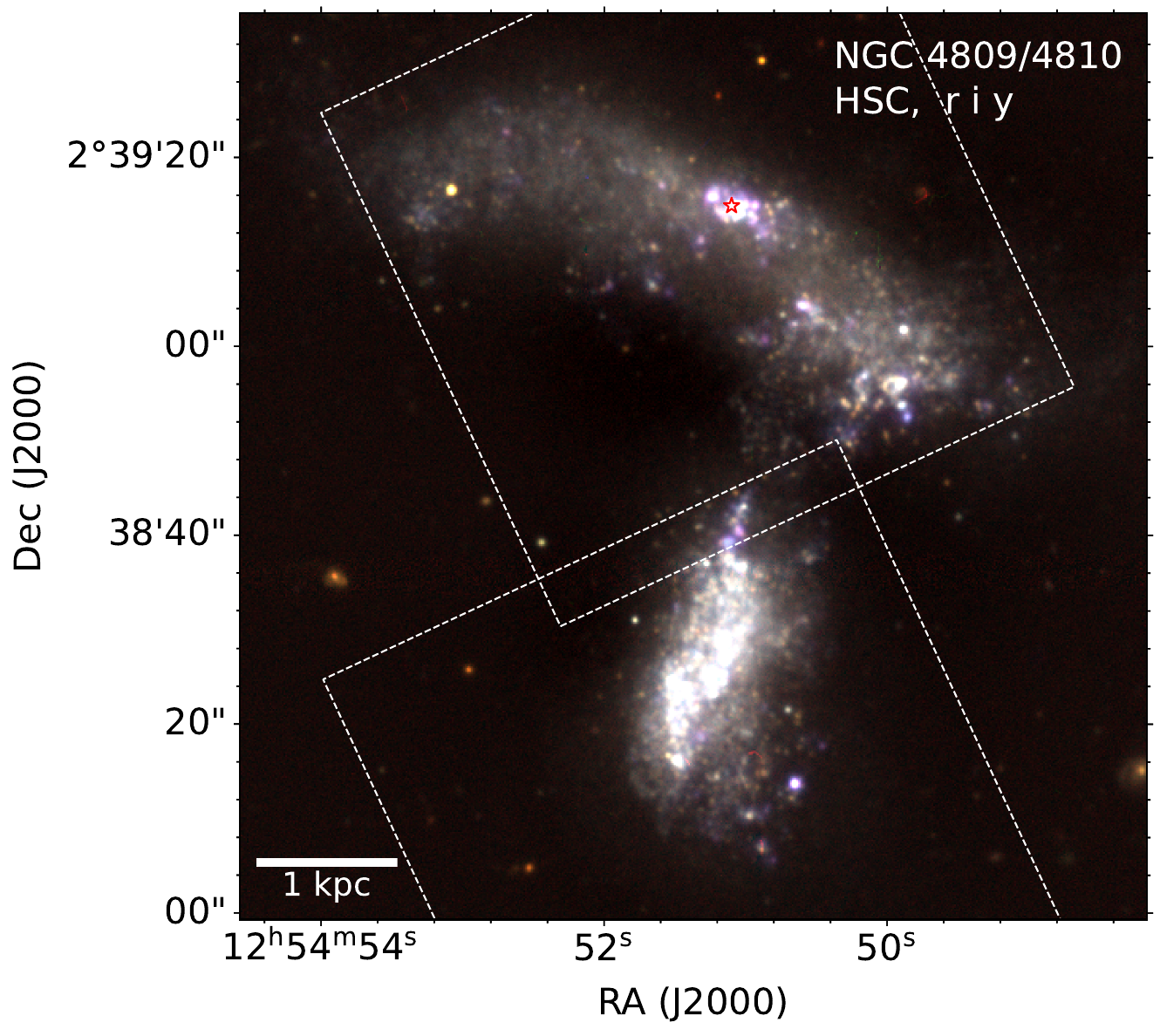}
   \includegraphics[width=0.45\textwidth]{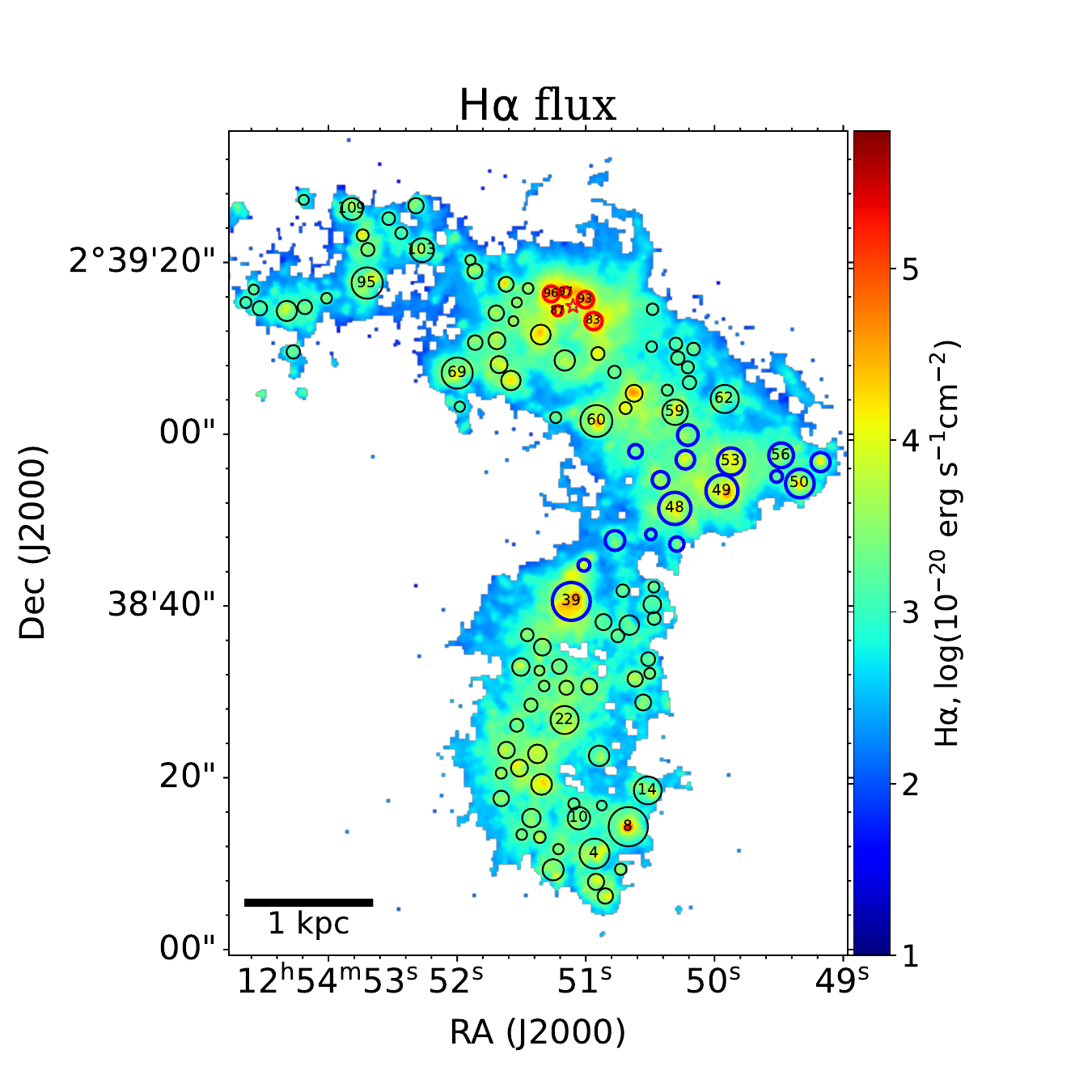}
   \caption{$\mathbf{Left:}$ A pseudo-color image of NGC 4809 (north component) and NGC 4810 (south component) combined with the $r$, $i$, and $y$ band images from the Hyper Suprime-Cam Subaru Strategic Program (HSC-SSP). The white rectangles indicate the two field-of-view (FoV) regions observed by MUSE. The red star represents the supernova SN2011jm. $\mathbf{Right:}$ An integrated intensity map of the attenuation-corrected $\ha$ emission. {The circles represent the identified $\ha$ knots detected by \textit{Astrodendro}. Red circles represent these knots around SN2011jm and blue circles mean these knots at the interaction area of two galaxy components.} We also label the IDs of some large knots.}
   \label{fig:Ha}
 \end{figure*}

We use the \textit{Astrodendro}\footnote{\url{https://dendrograms.readthedocs.io/en/stable/}} Python package \citep{Goodman2009Jan} to search for star-forming clumps in the $\ha$ map of NGC 4809/4810. This clump-finding algorithm is based on the dendrogram and has been used to identify reliable star-forming cores in galaxies (see \cite{Li2020c} for a detailed comparison of different clump-finding packages). To define the boundaries of clump structures, we specify several parameters, including $min\_value$, $min\_delta$ and $min\_npix$. The $min\_value$ represents the minimum value in the field to be considered, $min\_delta$ represents the minimum significance of the structure to avoid including local maxima, and $min\_npix$ specifies the minimum number of pixels that a structure should contain. Adopting the smallest region radius as the seeing value, we obtain a $min\_npix$ of 26. {We select six different regions in the $\ha$ intensity map around the galaxy, and measure the root mean square (RMS) values in each region. Then we adopt the $min\_value$ and $min\_delta$ as the mean values of 5$\times$ RMS and 1 $\times$ RMS, respectively.} Ultimately, we identify 112 $\ha$ knots within these two galaxies, and their distribution is shown in the right panel of Fig. \ref{fig:Ha}. 

 \begin{figure*}[t]
   \centering
   \includegraphics[width=0.45\textwidth]{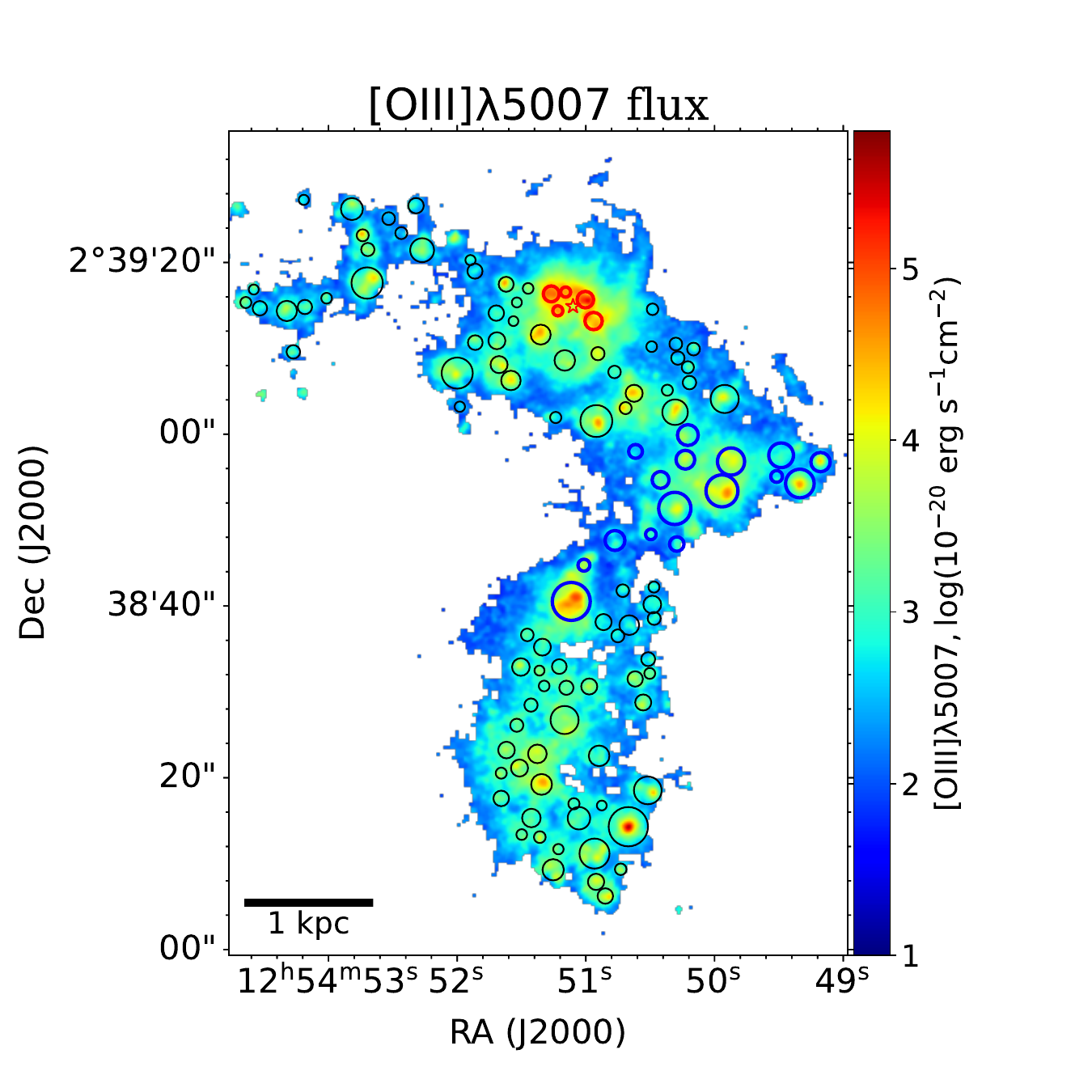}
   \includegraphics[width=0.45\textwidth]{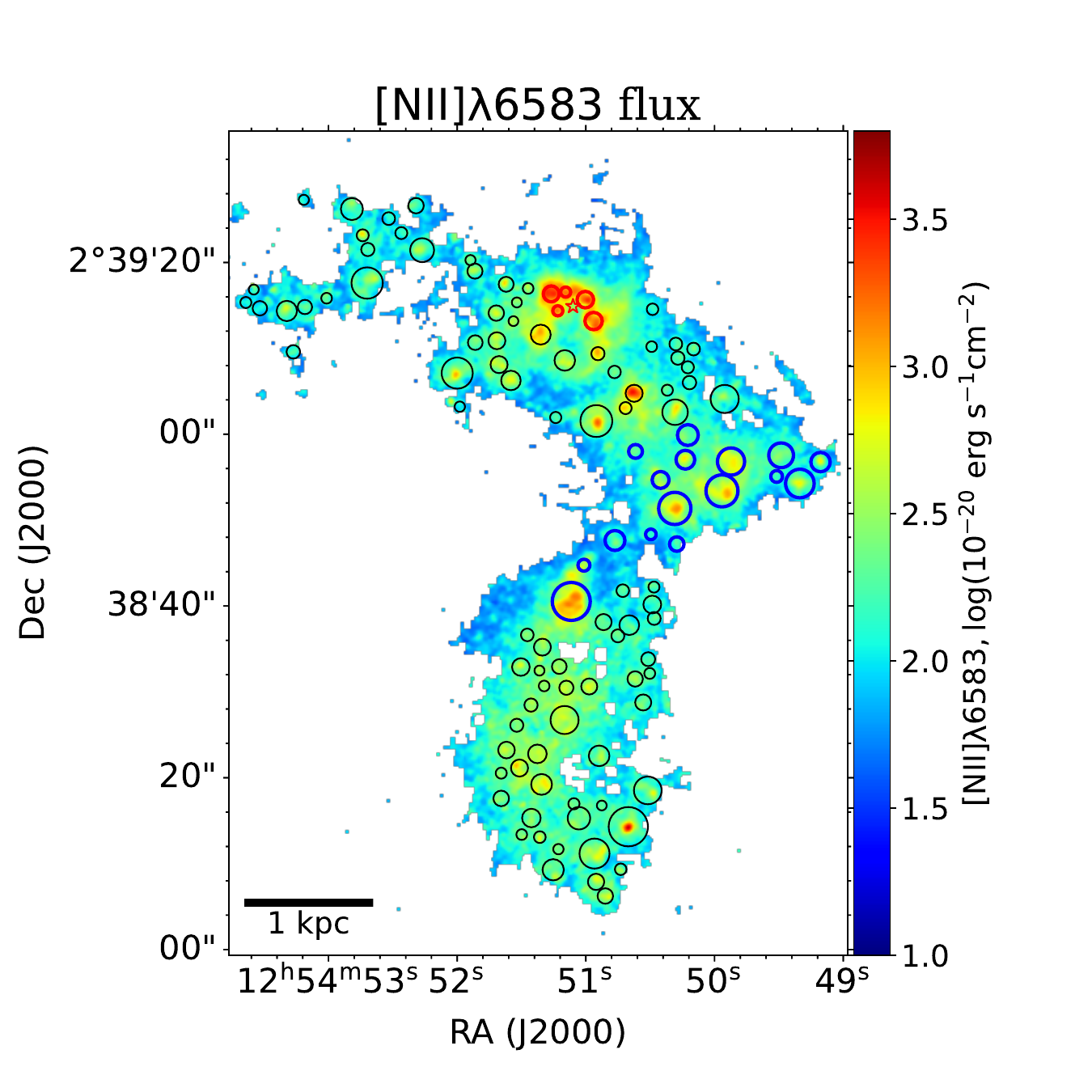}
   \includegraphics[width=0.45\textwidth]{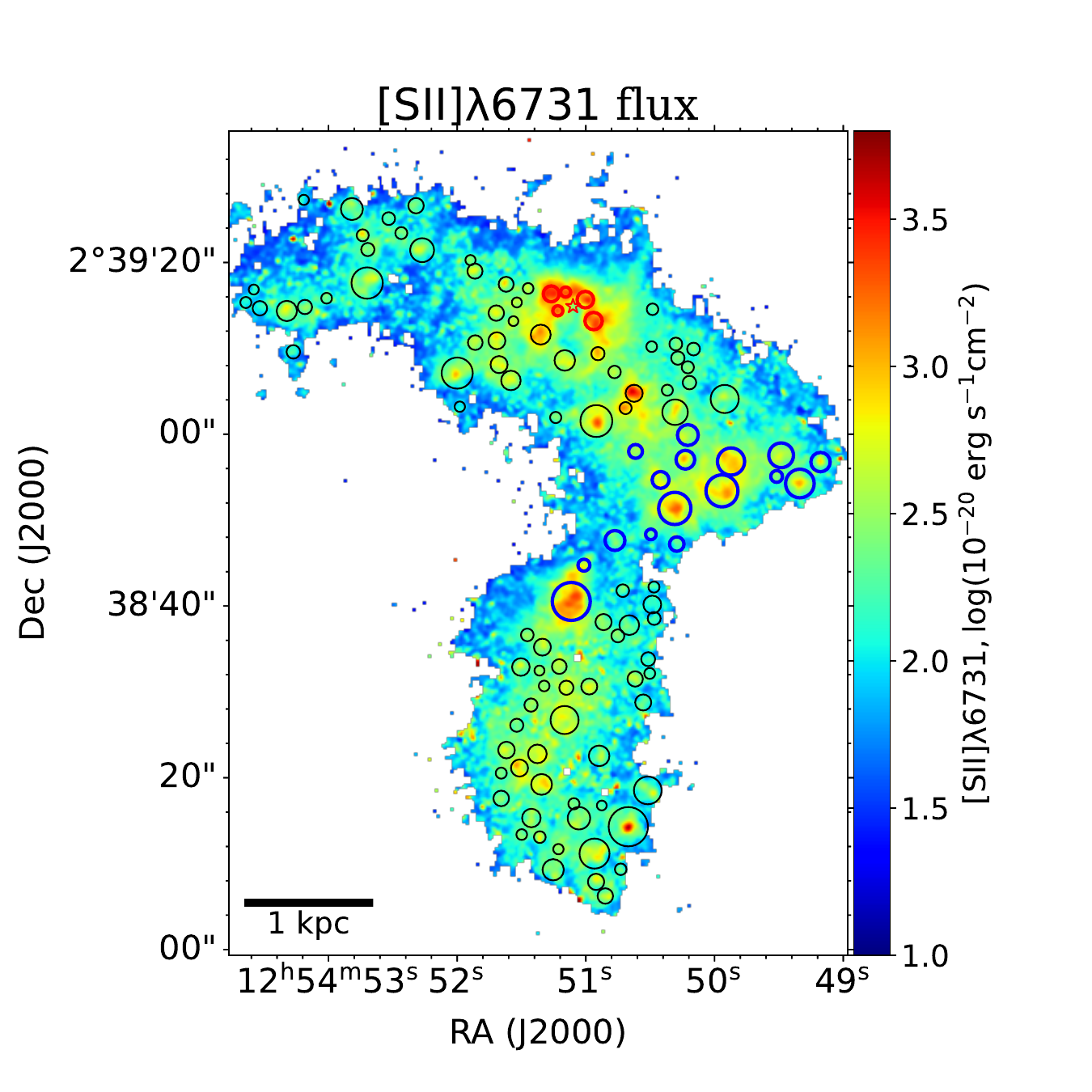}
   \includegraphics[width=0.45\textwidth]{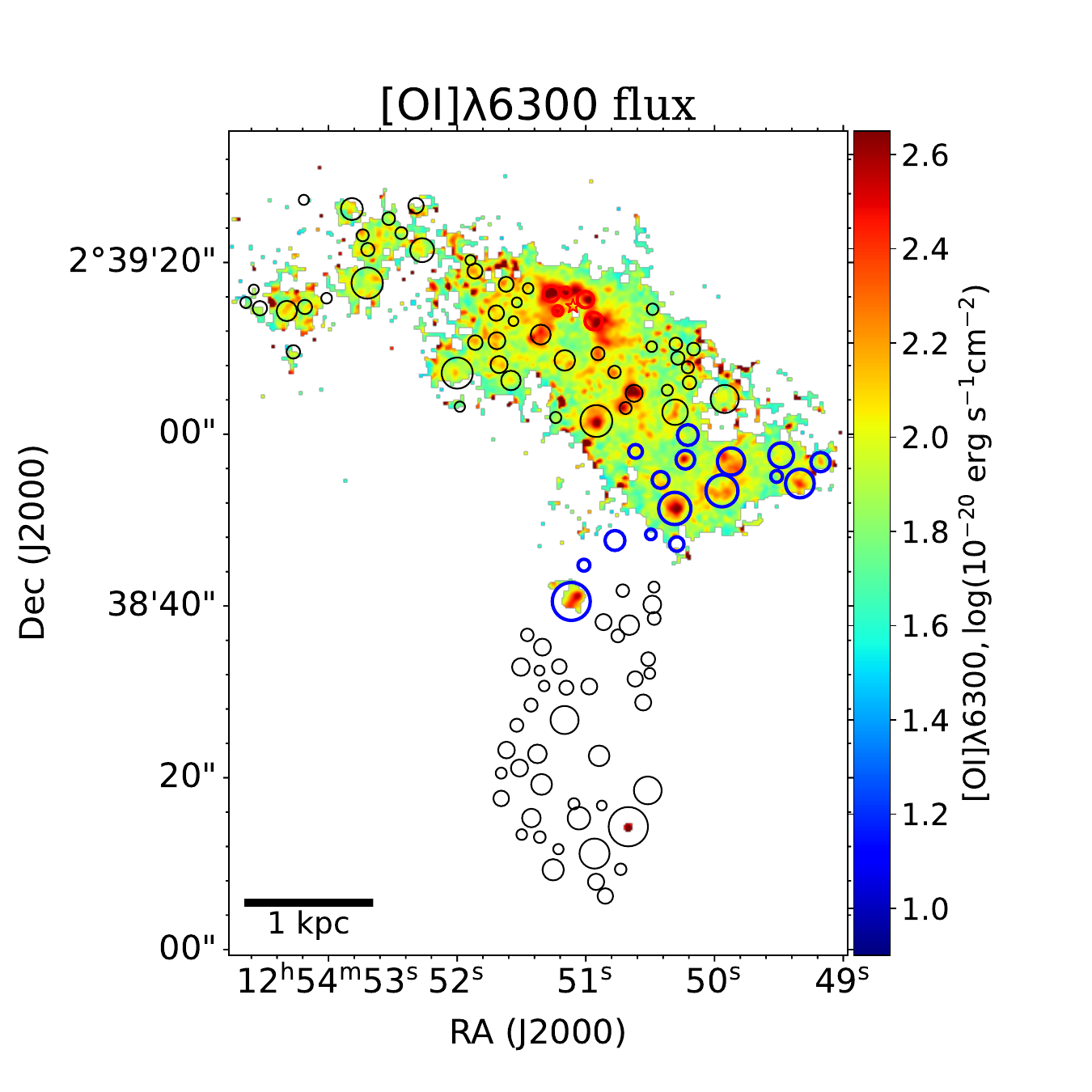}
   \caption{Integrated intensity maps of four emission lines, $\oiii\lambda5007$ (upper-left), $\nii\lambda6583$ (upper-right), $\sii\lambda6731$ (bottom-left), and $\oi\lambda6300$ (bottom-right), are shown after attenuation correction.  Circles indicate the locations of identified star formation knots, which are the same as those in Fig. \ref{fig:Ha}.}
   \label{fig:oiii_nii}
 \end{figure*}

 \begin{figure*}[t]
   \centering
   \includegraphics[width=0.45\textwidth]{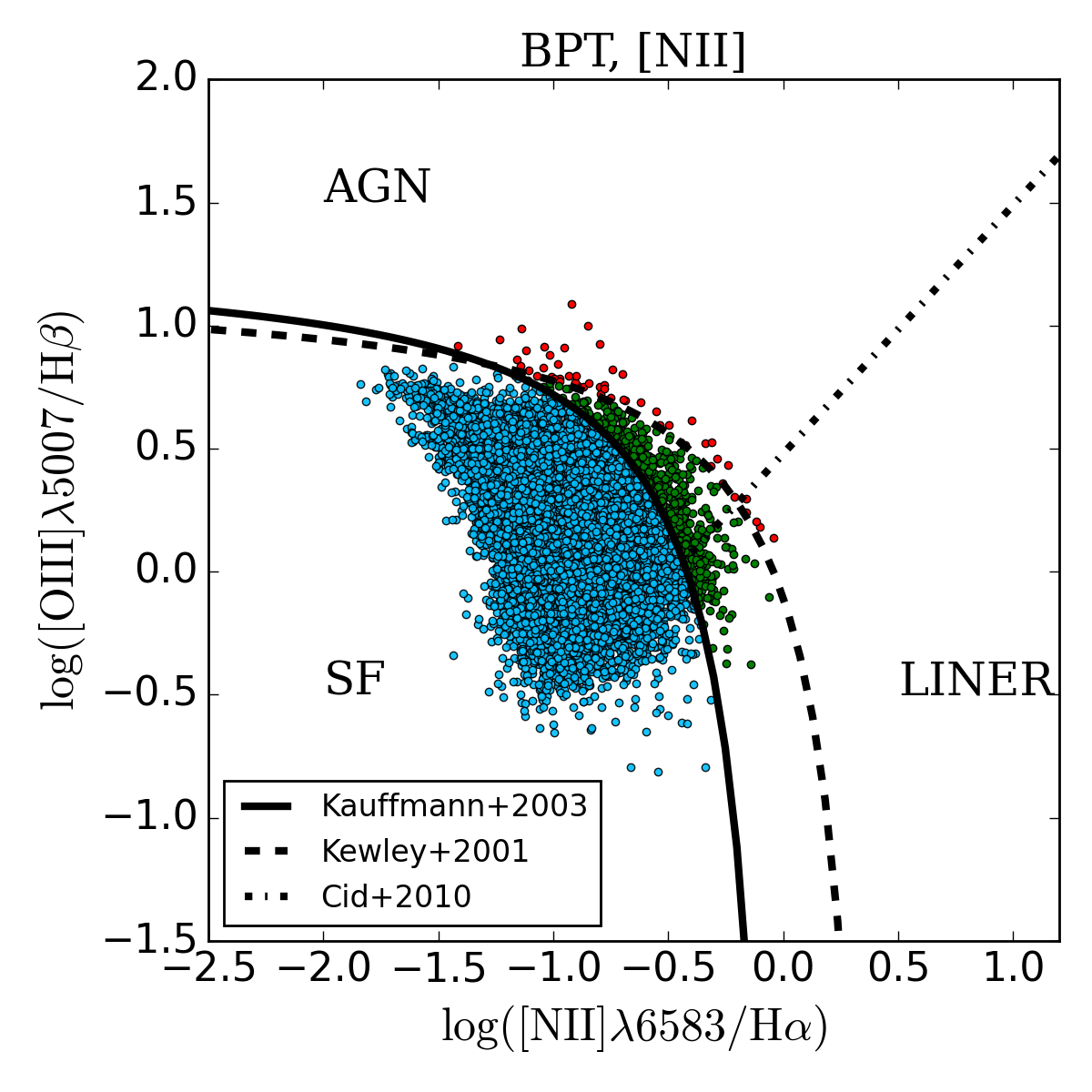}
   \includegraphics[width=0.45\textwidth]{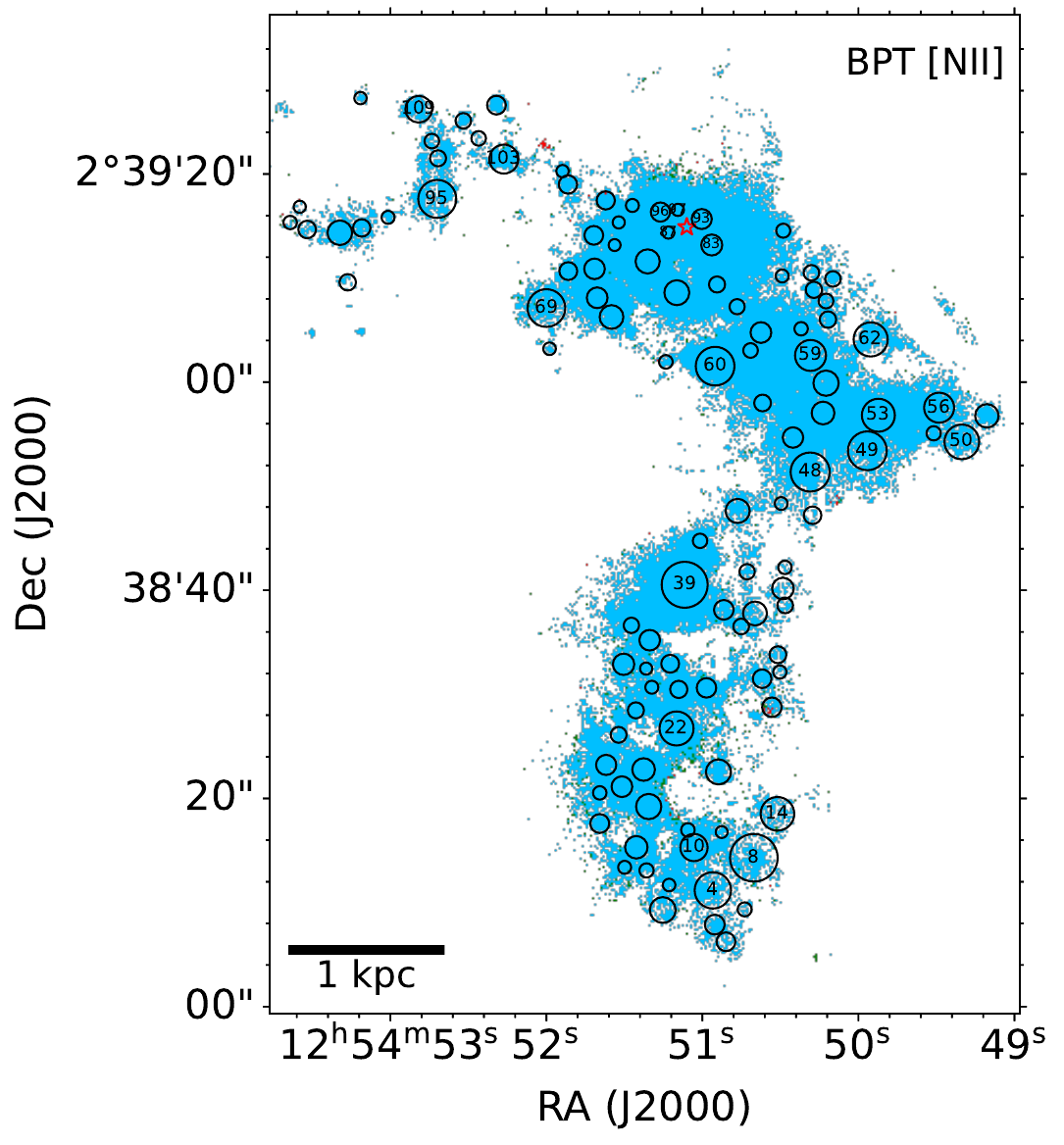}
   \includegraphics[width=0.45\textwidth]{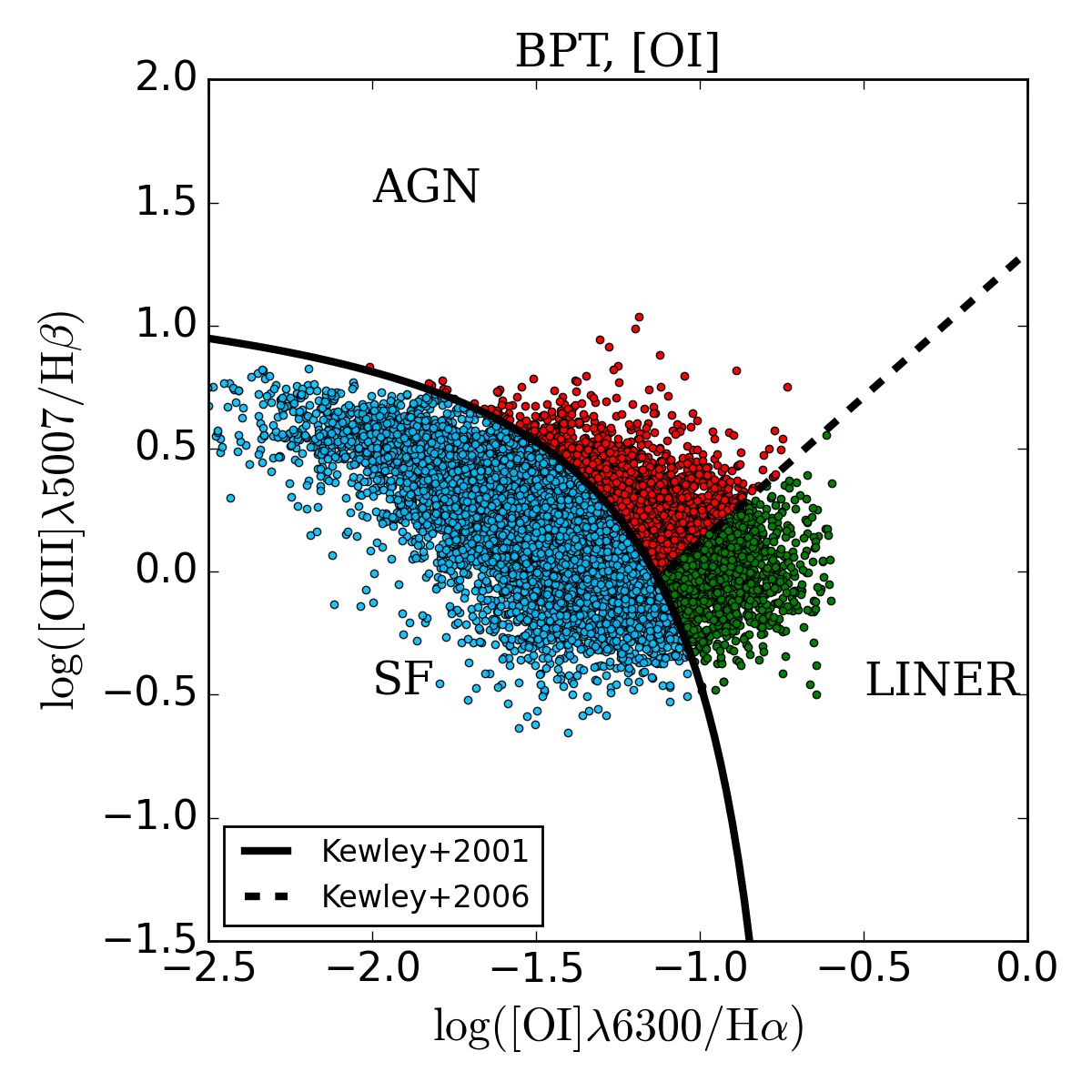}
   \includegraphics[width=0.45\textwidth]{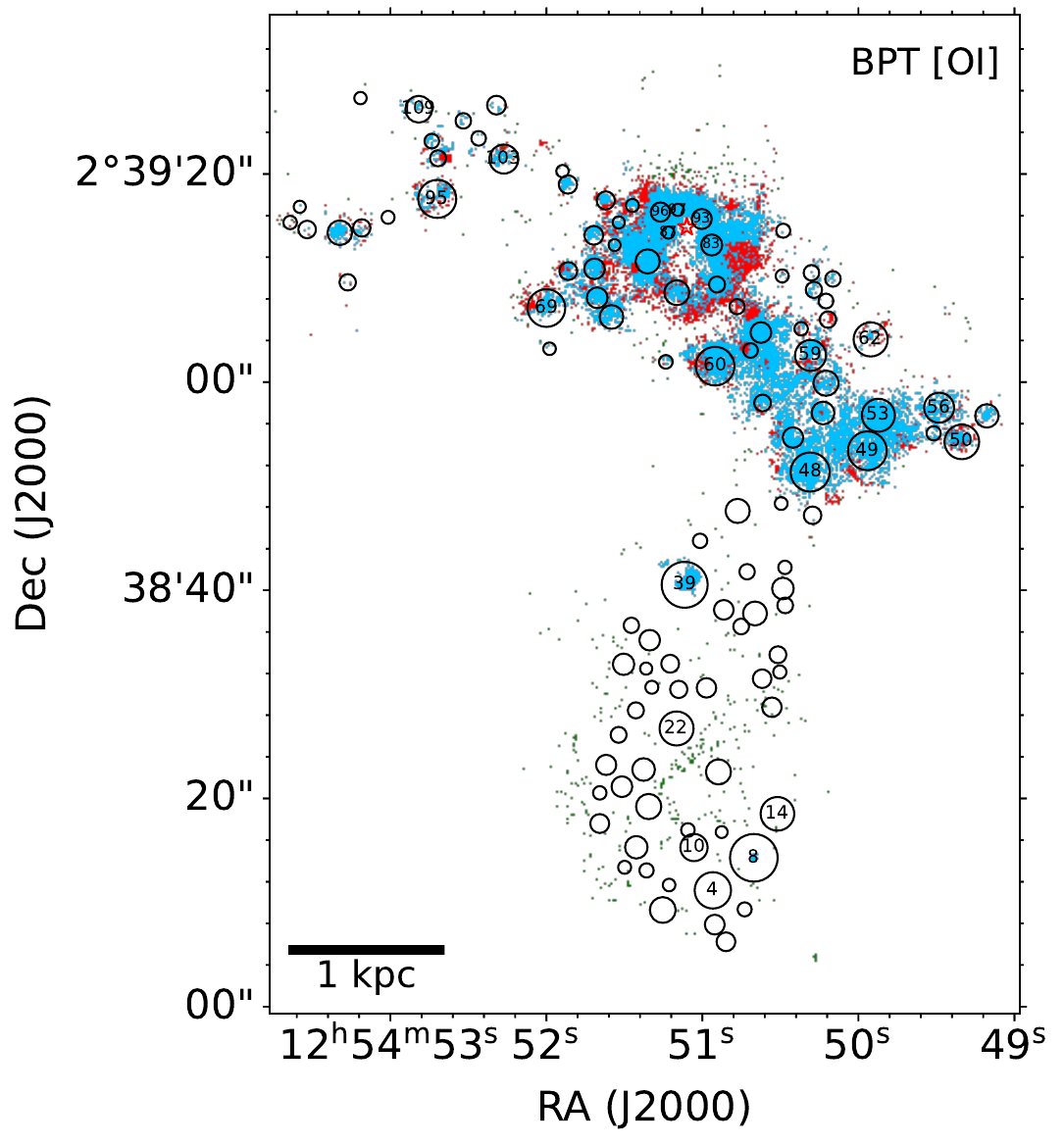}
   \caption{Spatially resolved BPT diagrams ($left$) of NGC 4809/4810 and their corresponding spatial distribution ($right$), obtained from the attenuation-corrected integrated intensity maps. The solid, dashed, and dotted lines in the $\nii$-based BPT diagram are the demarcation curves that separate star formation, AGNs, and Low Ionization Nuclear Emission Regions (LINERs) as defined by \cite{Kauffmann2003}, \cite{Kewley2001}, and \cite{cidfernandes2010}, respectively. In the $\oi$-based BPT diagram, solid and dashed lines separate star-forming galaxies, Seyfert galaxies, and LINERs as defined by \cite{Kewley2001} and \cite{kewley2006}. Regions dominated by star formation, AGN-type ionization, and LINERs are marked in blue, red, and green, respectively.}
   \label{fig:bpt}
 \end{figure*}

\section{Results}
\label{sec:resu}

The spatial distribution of ionized gas can provide crucial insights into the kinematics of gas and the activity of star formation. In this study, we present, for the first time, the properties of ionized gas within NGC 4809/4810.

\subsection{Emission line maps and BPT diagrams}
\label{subsec:bpt}

We show in Figures \ref{fig:Ha} and \ref{fig:oiii_nii} the emission line maps of the ionized gas, including $\ha$, $\oiii\lambda5007$, $\nii\lambda6563$, $\sii\lambda6731$, and $\oi\lambda6300$. These maps provide insight into the level of ionization in $\hii$ regions and the strength of the ultraviolet radiation field. For instance, as the ionization potential of H is much lower (13.6 eV) than that of O$^{++}$ (35 eV), higher $\oiii\lambda5007/\ha$ ratios indicate a higher number of energetic photons. The spatial distributions of $\oiii\lambda5007$ and $\ha$ emissions overlap, tracing the same ionizing source of young stars in star-forming regions. The luminous $\ha$ and $\oiii\lambda5007$ emission lines are mainly concentrated in the hotspots surrounding the supernova SN2011jm, which was regarded as a type-Ic supernova with an initial star mass of 20.7 $\msun$ \citep{howerton2011a,kuncarayakti2018a}, and the interaction regions of these two galaxies. Additionally, there are some isolated $\ha$ and $\oiii$-emitting blobs between these two galaxies. The spatial distributions of $\nii\lambda6583$ and $\sii\lambda6731$ are similar to $\ha$ and $\oiii$ emission and can be used to calibrate the gas-phase metallicity.

In addition to highly ionized $\oiii$ gas, the lowly ionized forbidden $\oi\lambda6300$ line is commonly used to trace the radiation of the ISM heated by shocks. Thermal electrons with low energy ($\sim$1.9 eV) can excite O atoms through collisions, resulting in the observed $\oi\lambda6300$ line. Higher speed shocks can increase the thermal energy of free electrons in the ISM, thereby enhancing the number of collisional excitations. In Fig. \ref{fig:oiii_nii}, the distribution of $\oi\lambda6300$ is similar in NGC 4809 but is only detected in two bright star formation knots of ID 8 and 39 in NGC 4810. This indicates that most shocks are occurring in NGC 4809 instead of NGC 4810, which will be discussed in Section \ref{subsec:kine_gas}.

We utilize spatially resolved BPT diagnostic diagrams \citep{Baldwin1981,Kewley2001,Kauffmann2003} to determine the primary ionizing source in each spaxel. The $\nii$-based \citep{Kewley2001,Kauffmann2003,cidfernandes2010} and $\oi$-based \citep{Kewley2001,kewley2006} BPT diagrams and their corresponding spatial distributions are presented in Fig. \ref{fig:bpt}. Our results reveal that star formation dominates gas ionization in both galaxies in the $\nii$-based diagram. However, in the $\oi$-based BPT diagram, we find a few AGN-like ionized regions located at the outer edge of the supernova ring, which might be caused by shock. A few spaxels identified as LINER-like regions may be excited by shock or ionized by hot evolved (post-asymptotic giant branch) stars \citep{Belfiore2016a}. {In summary, we discover that 104 $\ha$ knots are primarily determined by star formation in both the $\nii$-based and and $\oi$-based BPT diagrams, while 8 knots are detected at the AGN or LINER regimes of the $\oi$-based BPT diagram.}

\subsection{Kinematics of ionized gas}
\label{subsec:kine_gas}

\begin{figure*}[t]
   \centering
   \includegraphics[width=0.45\textwidth]{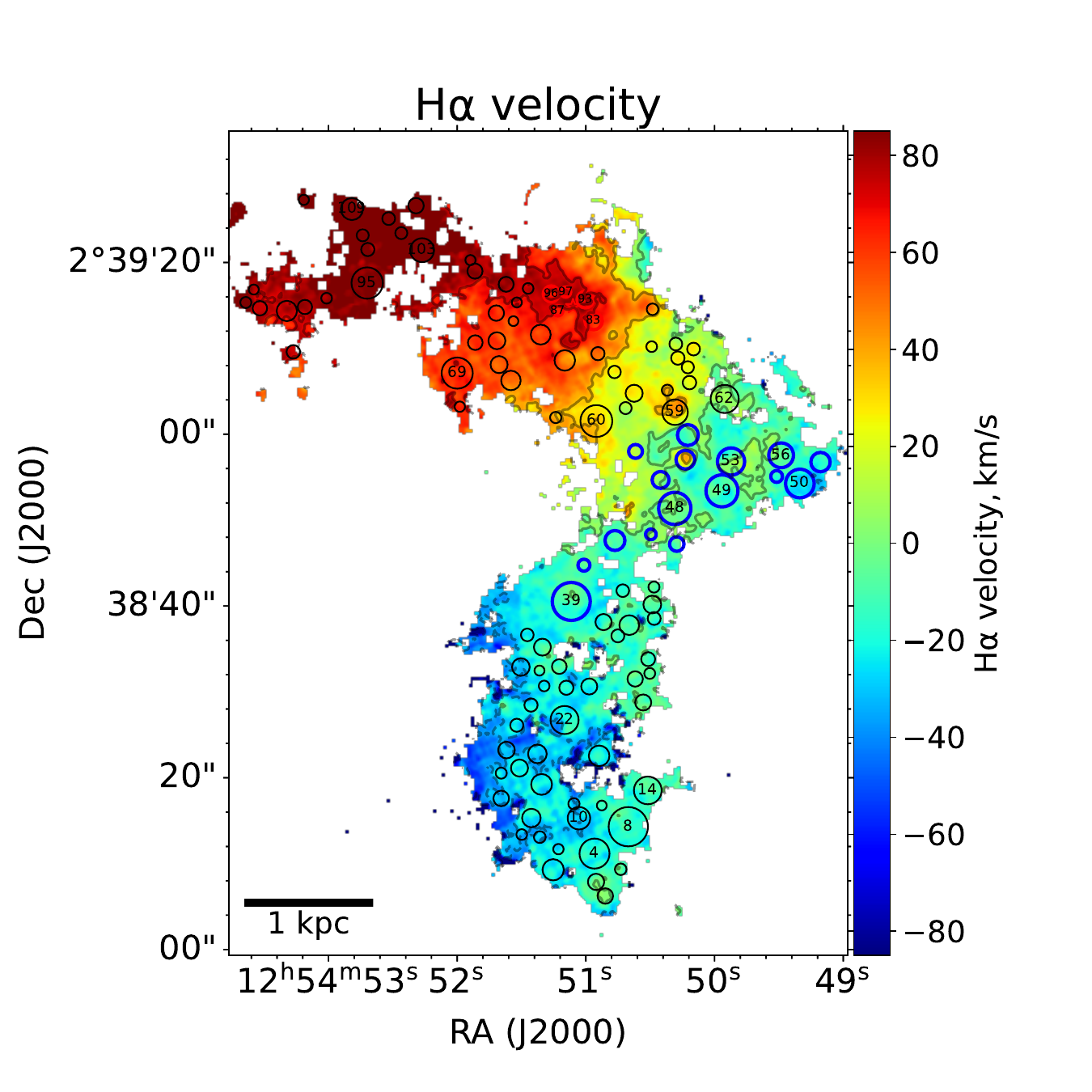}
   \includegraphics[width=0.45\textwidth]{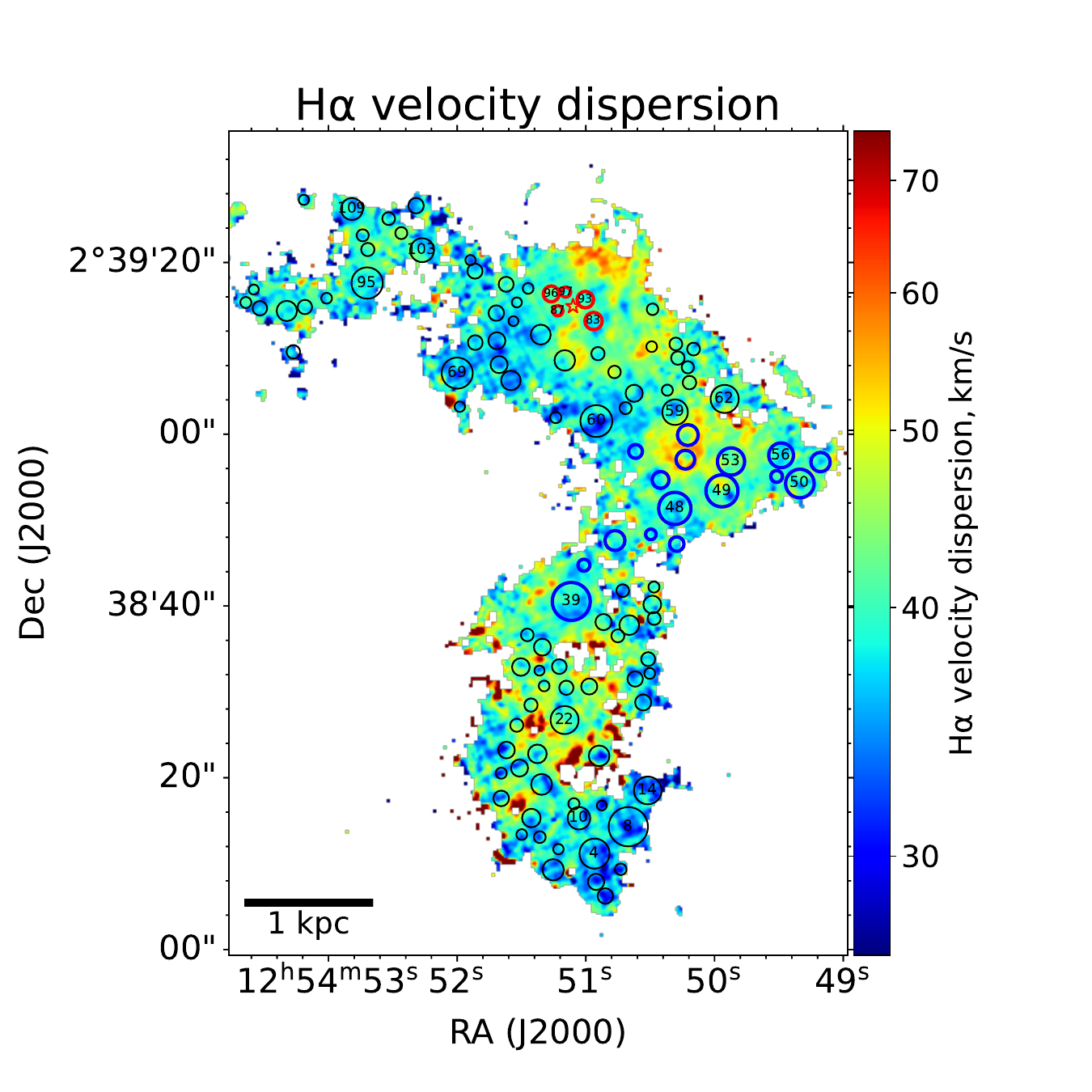}
   \includegraphics[width=0.45\textwidth]{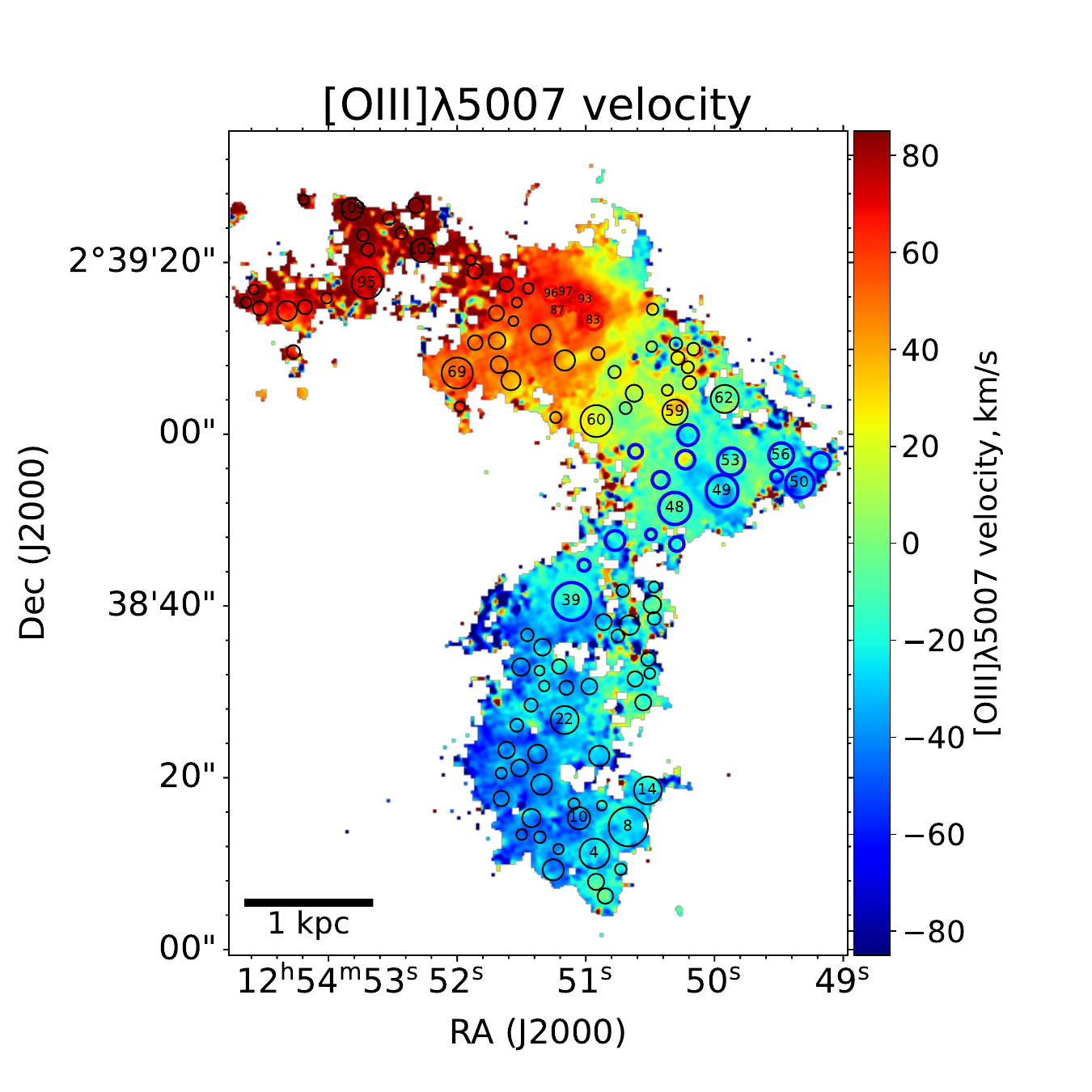}
   \includegraphics[width=0.45\textwidth]{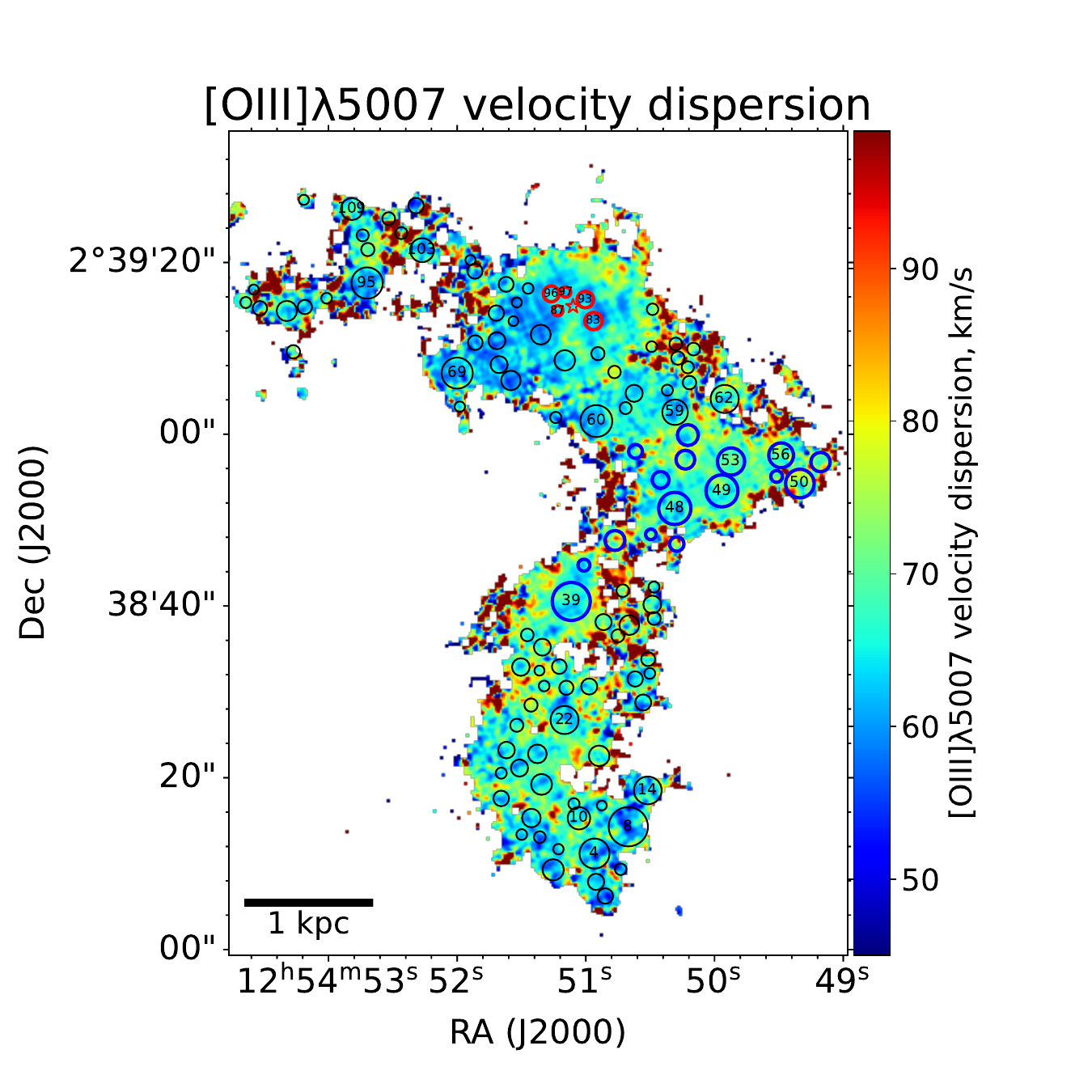}
    \caption{Velocity ($left$ panel) and velocity dispersion ($right$ panel) maps of the $\ha$ and $\oiii\lambda5007$ emission lines, respectively. The velocity dispersion is obtained by enforcing single Gaussian profile fitting. The gray contours in the $\ha$ velocity map represent the velocity levels of [-70, -35, 0, 35, 70] $\kms$.}
   \label{fig:ha_oiii_vel}
 \end{figure*}

We adopt a single Gaussian profile to fit the mean velocity of gas in NGC 4809/4810. In Fig. \ref{fig:ha_oiii_vel}, we present the velocity and velocity dispersion maps of the ionized gas, traced by $\ha$ and $\oiii\lambda5007$. The $\ha$ velocity field is overlaid as contours with levels of [-70, -35, 0, 35, 70] $\kms$. As seen in these maps, the ionized gas displays an obvious northeast-southwest velocity gradient in NGC 4809, while it is approximately oriented along the northwest-southeast direction in NGC 4810. We observe that the ionized gas in NGC 4809 is clearly separated into different velocity ranges, while it appears to be mixed in NGC 4810. 
Moreover, the southwest regions of NGC 4809 (e.g., knots ID: 48, 49, 50, 53, 56) exhibit similar velocities to those in NGC 4810 (knot 39). {This result suggests that these star-forming knots (IDs: 39, 48, 49, 50, 53, 56; marked with blue edgecolors) are located at the interaction area of the two merging galaxies.}

We compute the velocity dispersion ($\sigma$) using the FWHM ($\sigma$ = FWHM/2.355) and subtract the instrumental broadening of MUSE \citep{Bacon2010}. In the $right$ panels of Fig. \ref{fig:ha_oiii_vel}, we present the luminosity-weighted average velocity dispersion maps of the $\ha$ and $\oiii\lambda5007$ emission gas, which reflect the typical random motions of the ionized gas in the interstellar medium. The $\ha$ and $\oiii\lambda5007$ velocity dispersion range from 25 to 64 and 42 to 114 $\kms$, respectively, with median values of 40 and 66 $\kms$. Low dispersion values ($\sigma_{\ha} \leq 40 \ \kms$, $\sigma_{\oiii} \leq 65 \ \kms$) are measured at the northeast knots of NGC 4809 and south knots of NGC 4810, while larger dispersion values ($\sigma_{\ha} \geq 40 \ \kms$, $\sigma_{\oiii} \geq 65 \ \kms$) are observed at the knots of the interaction area. 
Furthermore, high $\ha$ and $\oiii$ velocity dispersion values are measured at the north and south of the supernova and its surrounding regions, indicating the possible influence of the massive stars/clusters on the stellar feedback in the past evolutionary stage.

\subsection{Distribution of gas-phase metallicity}
\label{subsec:metal}

\begin{figure*}[t]
   \centering
   \includegraphics[width=0.45\textwidth]{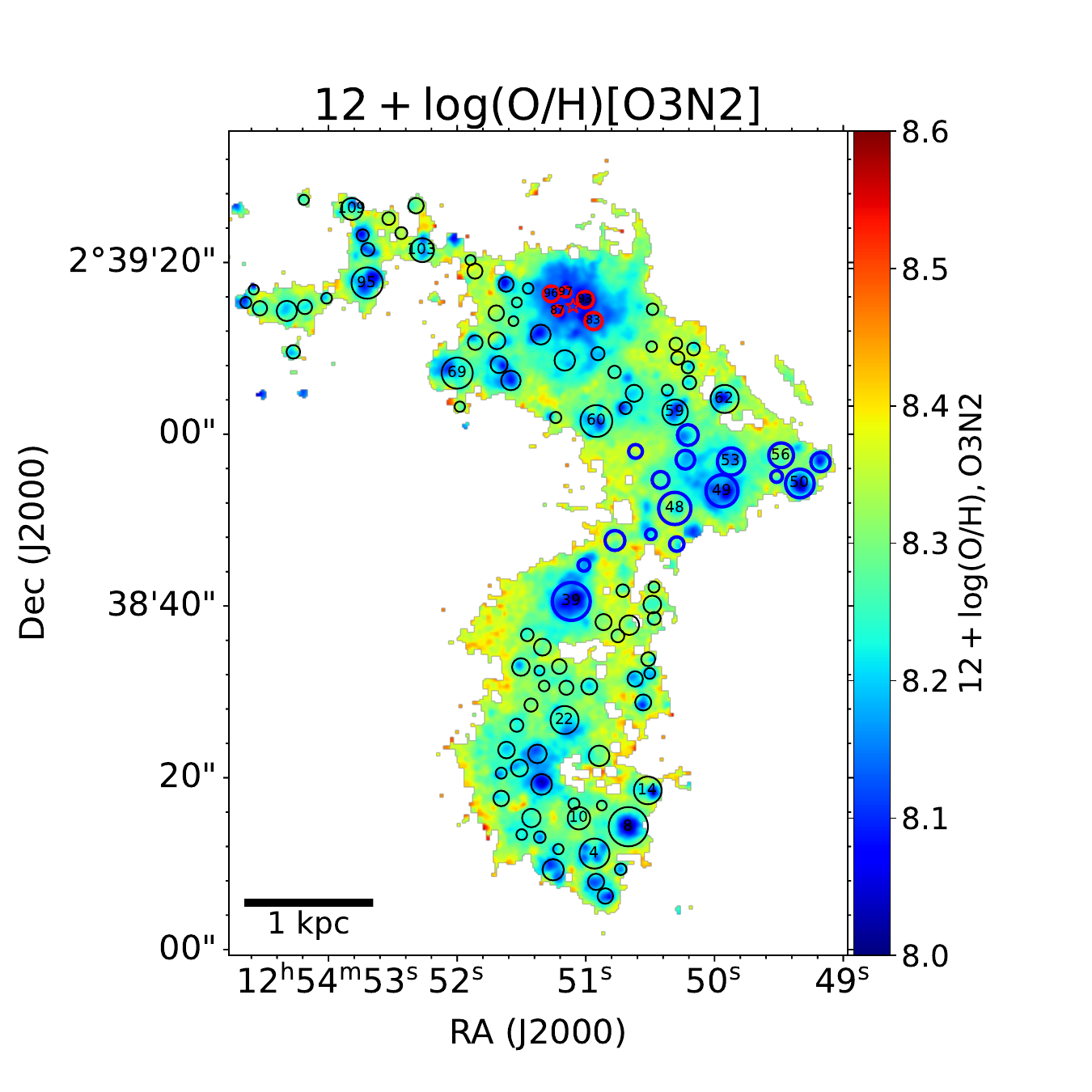}
   \includegraphics[width=0.45\textwidth]{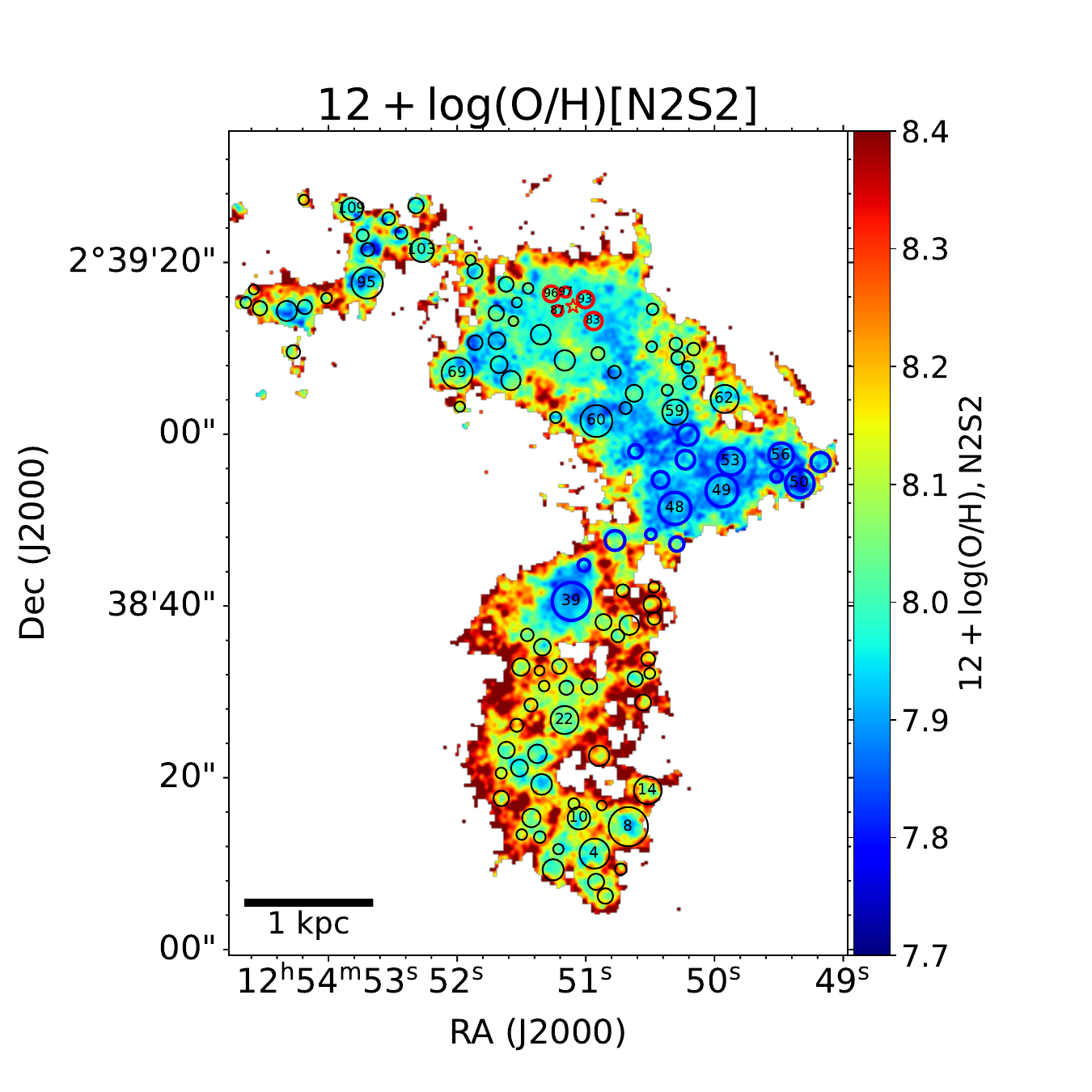}
   \caption{Gas-phase metallicity distributions of NGC 4809/4810 determined by two calibrators, O3N2 ($left$) and N2S2 ($right$), respectively.}
   \label{fig:zoh}
 \end{figure*}

The gas-phase metallicity is a crucial factor in understanding galaxy evolution, as it is linked to past star formation, mass inflow/outflow, and the mixing of metals in the ISM. There are various methods to measure the gas-phase metallicity, specifically the oxygen abundance, of the ISM \cite[e.g.,][]{kewley2008,kewley2019}. Some of these methods use photoionization models of $\hii$ regions to reproduce specific emission line ratios, such as N2O2 ($\nii\lambda6583/\oii\lambda3727$) \citep{kewley2002}, R23 ($(\oii\lambda3727+\oiii\lambda\lambda4959,5007)/\hb$) \citep{kobulnicky2004}, and N2S2 ($\nii\lambda6583/\sii\lambda\lambda6717,6731$) \citep{dopita2016}. The most reliable approach to determine metallicity involves directly measuring the electron temperature ($T_{\rm e}$) from faint auroral-to-nebular emission line ratios, such as $\oiii\lambda4363$/$\oiii\lambda5007$, and then using the strengths of other emission lines to calculate metallicity \cite[e.g.,][]{Izotov2006}. However, the faint $\oiii\lambda4363$ emission line is only significant in $\hii$ regions with high enough temperature and little cooling in metal-poor galaxies. Other calibrations use strong emission line ratios, such as O3N2 ($(\oiii\lambda5007/\hb)/(\nii\lambda6583/\ha)$) and N2 ($\nii\lambda6583/\ha$), to estimate the electron temperature metallicity. Since the MUSE spectral wavelength does not cover the $\oii\lambda3727$ and $\oiii\lambda4363$ lines, we use two different strong-line calibrators, O3N2 and N2S2, to determine the metallicity of each spaxel.

The O3N2 index \citep{Alloin1979} is defined as
\begin{equation}
\rm O3N2 \equiv \log\left(\frac{\oiii\lambda5007}{\hb} \times \frac{\ha}{\nii\lambda6583}\right).
\end{equation}
\cite{Marino2013} improved the O3N2 calibration based on CALIFA and literature data using $T_{\rm e}$, providing the following relation:
\begin{equation}
\rm 12 + \log(O/H) = 8.505 - 0.221 \times O3N2
\end{equation}
with a typical error of 0.08 dex when O3N2 ranges from $-1.1$ to 1.7. {The median O3N2-based metallicity and its 1$\sigma$ dispersion of all spaxels are about $8.25^{+0.07}_{-0.08}$.} 

The N2S2 index \citep{dopita2016} is defined as
\begin{equation}
\rm N2S2 \equiv \log\left(\frac{\nii\lambda6583}{\sii\lambda\lambda6717,6731}\right),
\end{equation}
which is sensitive to metallicity and weakly dependent on reddening and ionization parameter. Combining it with the N2 index ($\rm N2 \equiv \log(\nii\lambda6583/\ha)$), \cite{dopita2016} present the calibration relation as follows:
\begin{equation}
\rm 12 + \log(O/H) = 8.77 + N2S2 + 0.264 \times N2.
\end{equation}
{The median N2S2-based metallicity and its 1$\sigma$ dispersion are about $8.00^{+0.18}_{-0.12}$.} 

The gas-phase metallicity distributions are depicted in Figure \ref{fig:zoh}. {In this work, we regard the metallicities with lower (higher) than 1$\sigma$ value as metal-poor (rich), respectively.} It is evident that the metallicity distribution is heterogeneous across the two galaxies, comprising both metal-poor and metal-rich ionized gas. The majority of star formation knots exhibit deficient metallicity when using either the O3N2 or N2S2 calibrators, although the distributions vary considerably between the two.  In the O3N2-based map, metal-poor regions are primarily concentrated in the SN2011jm-surrounding regions (knots ID: 83, 87, 93, 96, 97; {marked with red edgecolors}), the interaction area ({marked with blue edgecolors}) between NGC 4809 and NGC 4810. However, in the N2S2-based map, metal-poor regions are mainly confined to the interaction area.

In summary, the star-forming knots at the interaction area show lower metallicity, regardless of the assumed calibrators. This result could be explained by the dilution of inflowing metal-poor gas during the merging process of the two dwarf galaxies. However, the metallicity of the regions surrounding SN2011jm differs significantly when using the two calibrators. The most deficient O3N2-based metallicity values around the supernova could be attributed to the calibration underestimation in this high ionization environment. If we use the N2S2 index, which is weakly dependent on ionization parameter, the ISM around the supernova exhibits high metallicity, possibly due to the substantial contribution of metal-enriched stellar wind (and supernova ejecta) from previous massive star clusters.

\subsection{Spatially-resolved star formation rate}
\label{subsec:sfr}

\begin{figure*}[t]
   \centering
   \includegraphics[width=0.45\textwidth]{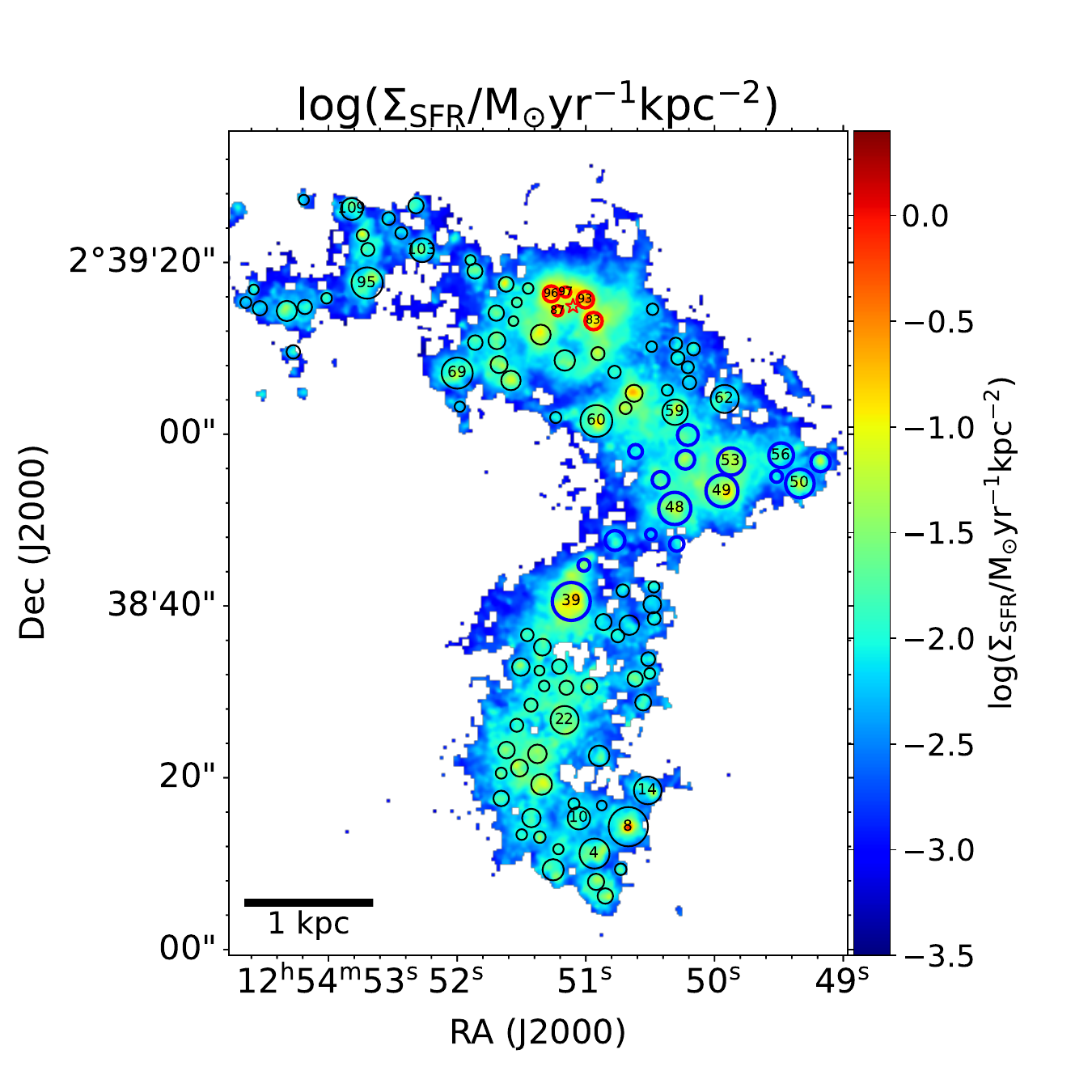}
   \includegraphics[width=0.45\textwidth]{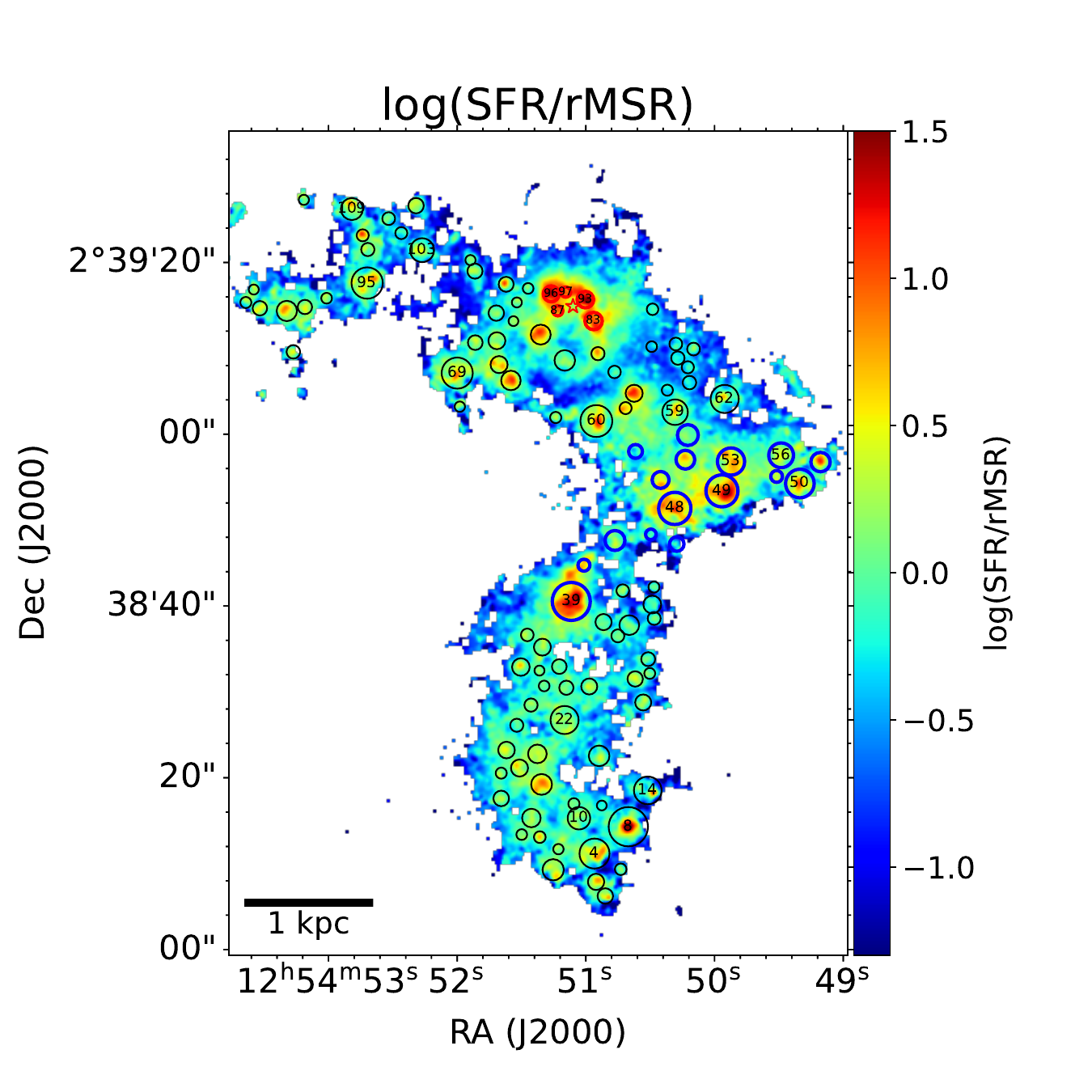}
   \caption{Maps of SFR surface density ($\Sigma_{\rm SFR}$, $left$ panel) and the ratios ($right$ panel) of $\Sigma_{\rm SFR}$ and the spatial resolved main sequence relation in nearby star forming galaxies from MaNGA survey \citep[rMSR,][]{liu2018}. }
   \label{fig:delta_msr}
 \end{figure*}

We calculate the extinction-corrected $\ha$ flux using the $\ha/\hb$ ratio and the ``Case B'' recombination model with the reddening formalism from \cite{Calzetti2000a}. Since our focus is on the intense star-forming regions in NGC 4809/4810, we use ionization-based tracers such as $\ha$ for this study. We obtain the SFR from the $\ha$ luminosity, $L_{\rm cor}(\ha)$, using the calibration relation in \cite{Kennicutt1998}, which assumes a \cite{Chabrier2003} IMF:
\begin{equation}
{\rm SFR}(\msun {\rm \ yr^{-1}}) = 4.4 \times 10^{-42} \times L_{\rm cor} \rm (\ha)(erg \ s^{-1}).
\end{equation}
The uncertainty in the SFR is calculated from the $\ha$ luminosity and attenuation uncertainties, which includes the errors in the $\hb$ intensity. The total SFR of these two galaxies is 0.12 $\pm$ 0.01 $\msun \rm \ yr^{-1}$. We also obtain the SFR surface density, $\Sigma_{\rm SFR}$, by dividing the area of each region. 

In the $left$ panel of Fig. \ref{fig:delta_msr}, we show the distribution map of the SFR surface density ($\Sigma_{\rm SFR}$). To compare this with the SFR in normal star-forming galaxies, we present a ratio map ($right$ panel) of the $\Sigma_{\rm SFR}$ to the resolved stellar mass-SFR relation (rMSR) in star-forming galaxies from the MaNGA survey \citep{liu2018}. {\cite{liu2018} studied 141,114 spaxel bins of star forming main sequence galaxies, then derived the rMSR by performing the best optimized least square (OLS) linear fitting for all spaxel bins.} We find that the knots with the highest SFRs and SFR/rMSR ratios ($>$ 1 dex) are located at the interaction area and the ring around the supernova. These SFR-enhanced regions also exhibit deficient metallicity in Fig. \ref{fig:zoh}. Our results suggest that the merging process of two dwarf galaxies can trigger their star formation activities at the interaction area even in a metal-poor environment.

\section{Discussion}
\label{sec:dis}

\subsection{Chemical inhomogeneity}
\label{subsec:dis_metal}

In Section \ref{subsec:metal}, we observe that the spatial distributions of metallicity in NGC 4809/4810 are significantly inhomogeneous, which contrasts with the homogeneous distribution found in some studies of $\hii$ galaxies on spatial scales larger than 100 pc \cite[e.g.,][]{cairos2009,garcia-benito2012,kehrig2016a}. These studies focus on significantly compact galaxies with star formation activities mostly concentrated in a few knots, where feedback mechanisms such as stellar winds and supernovae efficiently mix newly-created metals with the surrounding ionized gas homogeneously during the $\hii$ region lifetime.

However, previous IFU-based studies \cite[e.g.,][]{lagos2009,bresolin2019,james2020} have also reported chemical inhomogeneities. These studies suggest that complex processes, such as metal-enriched gas outflows from supernovae, self-enrichment winds from massive stars, metal-poor gas accretion from interactions/mergers, and rapid starburst activities at shorter timescales than the metal mixing, may be responsible for these inhomogeneities. 

Fig. \ref{fig:zoh} shows that the gas-phase metallicity is generally lower in most of the star formation knots compared to the surrounding ISM. The metallicity distributions derived by the O3N2 and N2S2 diagnostics exhibit partial similarities and differences. For example, the interaction areas exhibit deficient metallicity in both calibration maps, which could be attributed to the accretion/dilution of external metal-poor gas during the merging phase. We observe significant differences in the metallicity distributions around the supernova, with lower O3N2-based metallicity and higher N2S2-based metallicity. This difference may be due to the large discrepancy between different metallicity calibrators \citep[e.g.,][]{kewley2008,dopita2016,Morisset2016}. \cite{kewley2002} used the $\oiii/\oii$ two-line ratio to define the ionization parameter. Higher ionization parameters are always directly linked to higher sSFRs in star-forming galaxies \citep{kaasinen2018} due to the larger reservoir of ionizing photons at intense star formation sites. Although the $\oii\lambda3727$ emission line is not covered in our spectra, these supernova-surrounding regions exhibit much higher SFRs and sSFRs (Fig. \ref{fig:msr_zoh}), suggesting high ionization parameters. The gas-phase metallicity derived from the O3N2 diagnostic is additionally dependent on the ionization parameter, tending to show lower metallicity with higher ionization parameter \cite[e.g.,][]{pettini2004,Marino2013,Teklu2020a}. However, the N2S2 diagnostic is weakly dependent on the ionization parameter \citep{dopita2016}. These calibration differences may explain the different metallicity behaviors at the star formation sites around the supernova. The detailed dispersal and mixing of the metals released by massive stars/supernovae into the surrounding $\hii$ regions remains an open question that we will explore in future work (Gao et al. in prep).

\subsection{Scaling relations}
\label{subsec:scale_relation}

\begin{figure*}[t]
   \centering
   \includegraphics[width=0.45\textwidth]{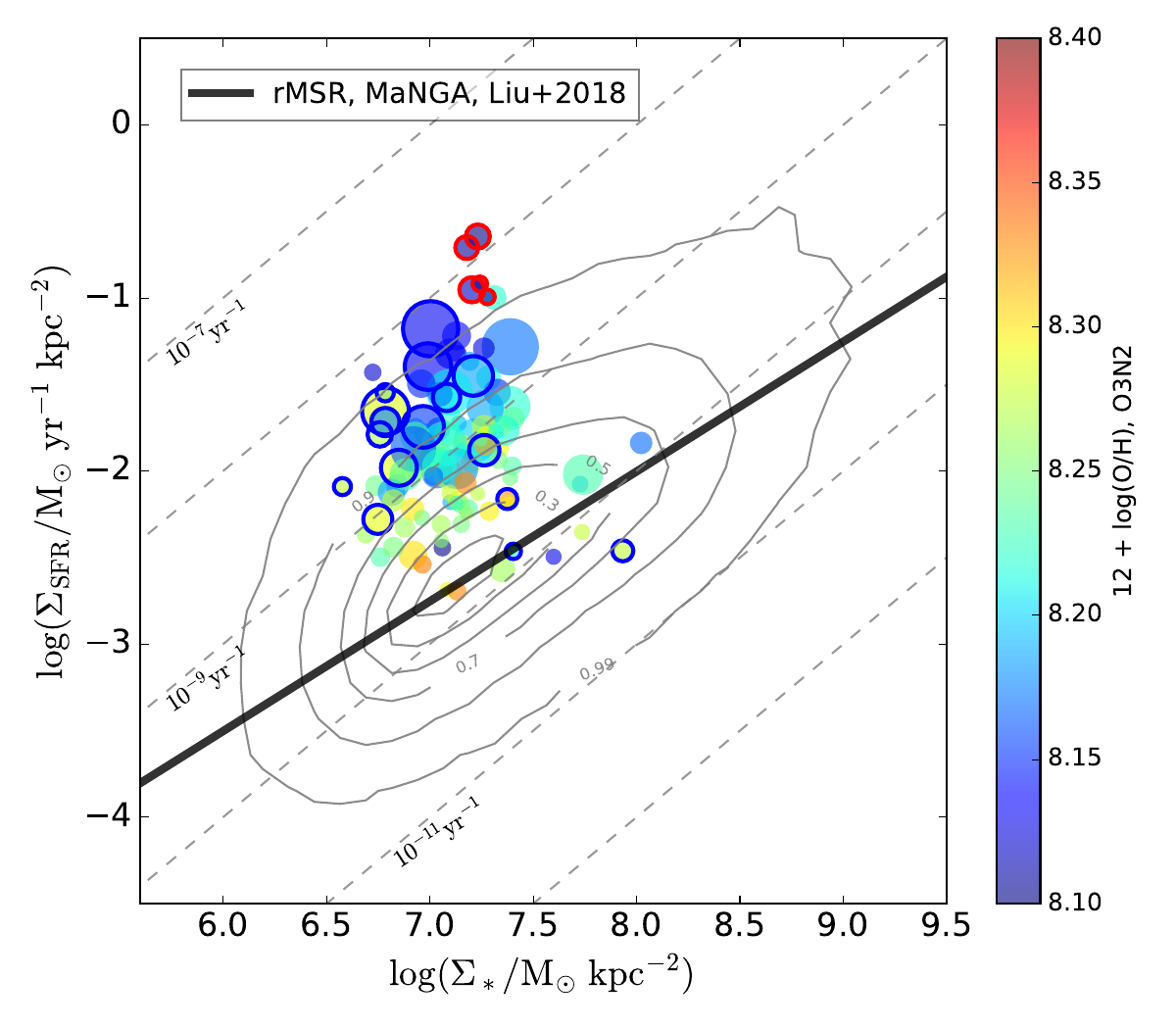}
   \includegraphics[width=0.45\textwidth]{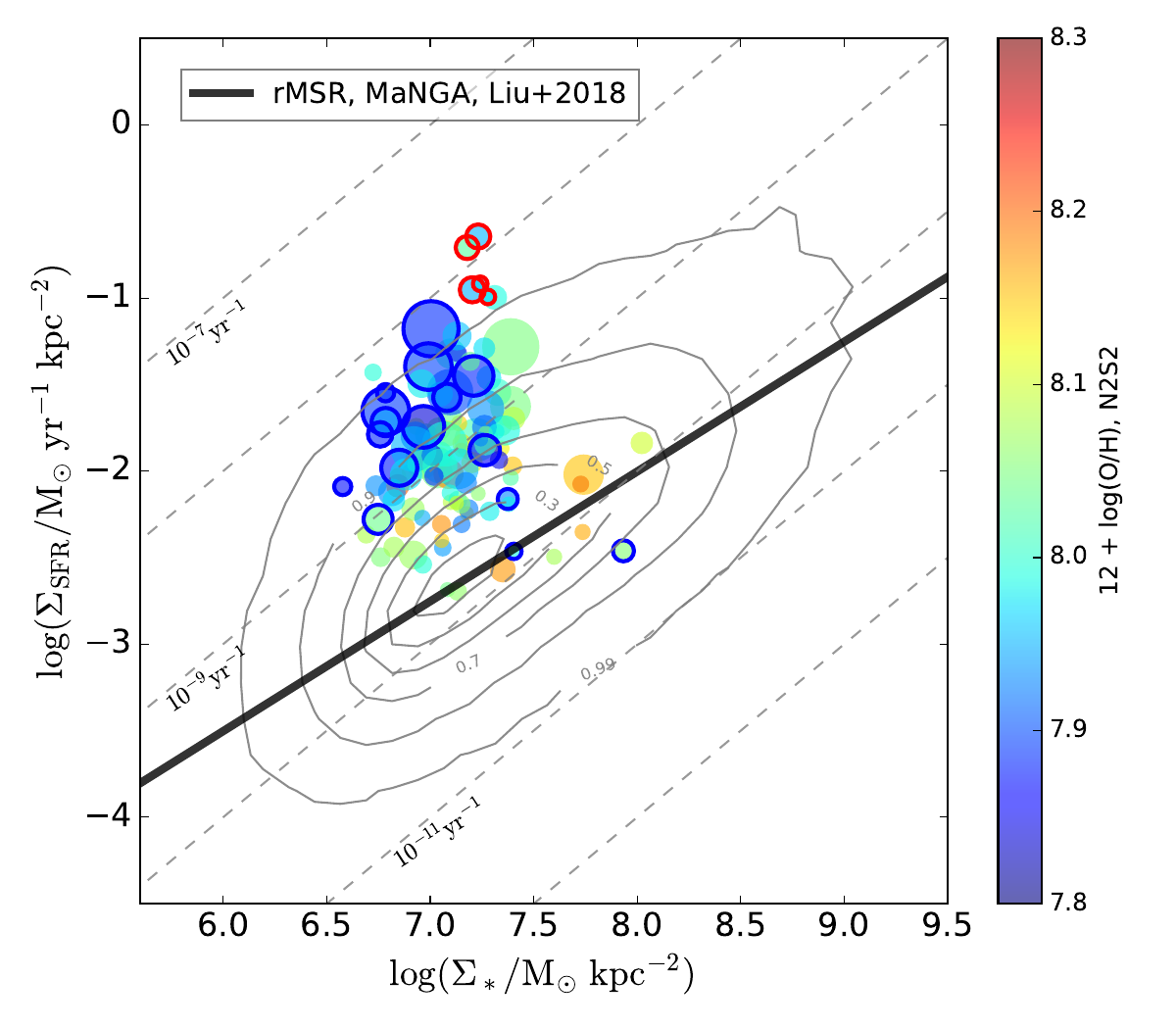}
   \caption{Spatially resolved main sequence relation (rMSR) between the surface density of stellar mass ($\Sigma_*$) and the surface density of star formation rate ($\Sigma_{\rm SFR}$) for star-forming knots in NGC 4809/4810. The sizes of the markers correspond to the sizes of the knots, while their colors indicate their gas-phase metallicity values determined by O3N2 (left panel) and N2S2 (right panel) diagnostics. The gray contours and black solid line represent the distribution of normal star-forming galaxies and their rMSR obtained by \cite{liu2018} from the MaNGA survey, respectively. {We also provide the star formation knots around SN2011jm (with red edgecolors) and at the interaction area (with blue edgecolors), which are the same as those in Fig. \ref{fig:Ha}.} The dashed lines represent different specific star formation rate (sSFR, $\Sigma_{\rm SFR} / \Sigma_*$) levels. }
   \label{fig:msr_zoh}
 \end{figure*}

In Section \ref{subsec:sfr}, we present the spatial distribution of SFR surface density and $\Sigma_{\rm SFR}$/rMSR. In this section, we analyze the star-forming knots in the $\Sigma_* - \Sigma_{\rm SFR}$ space and their gas-phase metallicity values, as shown in Fig. \ref{fig:msr_zoh}. The size of the markers indicates their knot sizes, and the colors indicate their gas-phase metallicity values determined by O3N2 and N2S2 diagnostics. We also mark the ID of some knots to facilitate the description of the star formation activities. In Fig. \ref{fig:msr_zoh}, the gray contours represent the $\Sigma_* - \Sigma_{\rm SFR}$ distribution of normal star forming galaxies from the MaNGA survey, and the solid line represents their best-fitted rMSR derived by \cite{liu2018}.  Dashed lines represent different specific SFR levels.

We find that most of the star-forming knots in our sample are located above the rMSR, indicating intense star formation activities. The highest sSFR values were found in knots surrounding supernovae ({marked as red edgecolors}), which were approximately two orders of magnitude higher than the rMSR. These values suggest that some physical mechanisms must be triggering star formation in these regions.
Furthermore, we note that knots located at the interaction areas between the two galaxies ({marked as blue edgecolors}) also exhibited sSFR values approximately one order of magnitude higher than the rMSR. These knots displayed low O3N2(N2S2)-based metallicity values and larger physical sizes, indicating that the merger process can trigger star formation activity on a large scale even in a metal-poor environment. 

\begin{figure*}[t]
   \centering
   \includegraphics[width=0.45\textwidth]{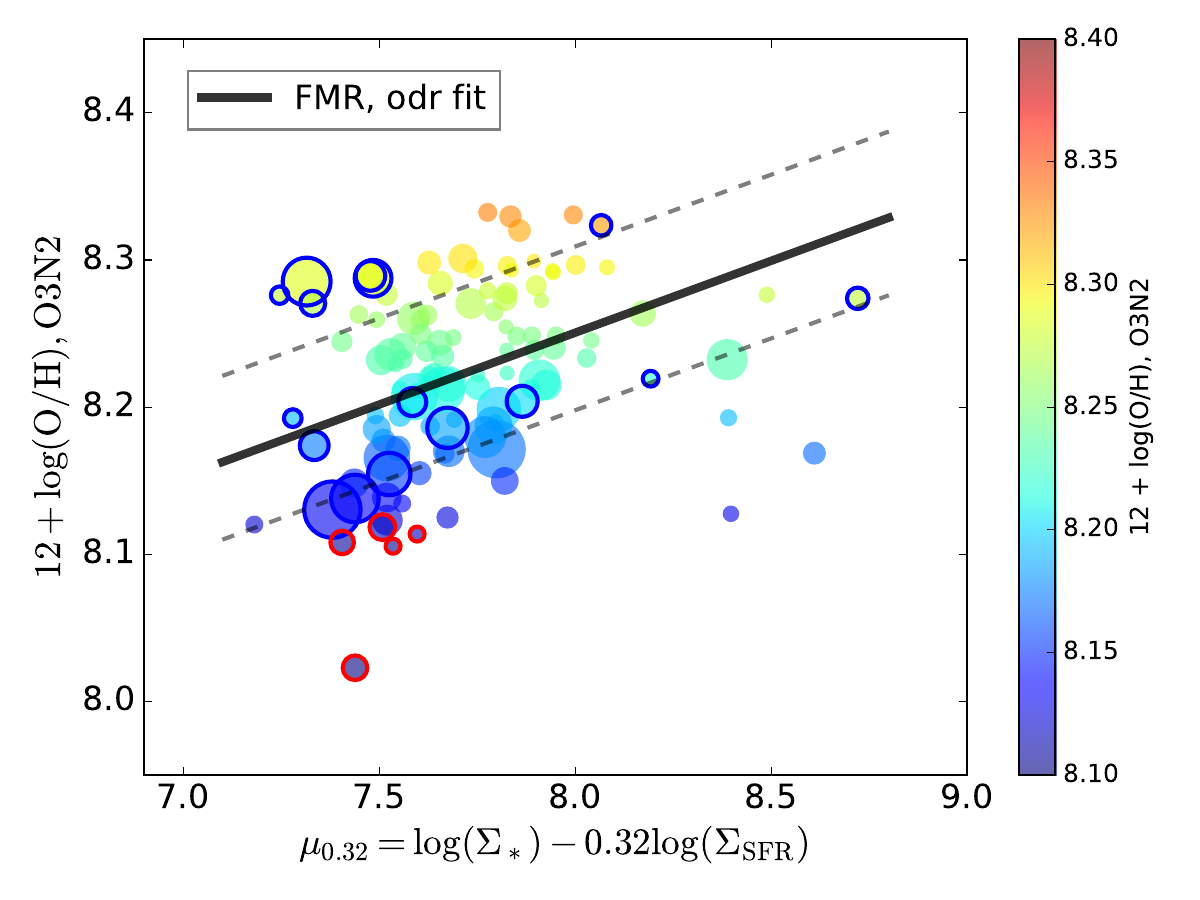}
   \includegraphics[width=0.45\textwidth]{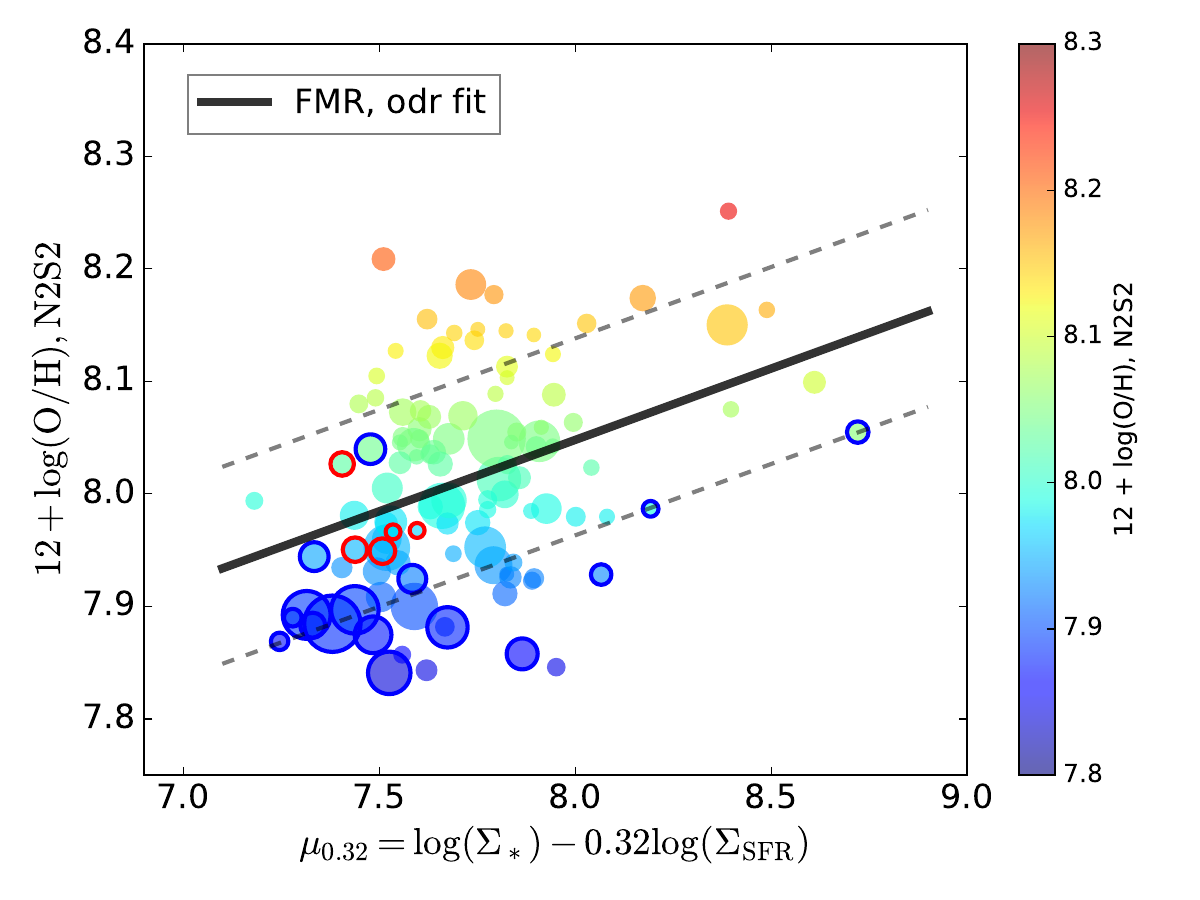}
   \caption{The Spatially resolved fundamental metallicity relation (rFMR, $\Sigma_* - Z - \Sigma_{\rm SFR}$) of star formation knots in NGC 4809/4810. The combined quantity $\mu_{\alpha}$ is defined as $\rm log(\Sigma_*) - \alpha log(\Sigma_{\rm SFR})$. We adopt the $\alpha = 0.32$, which is same to previous studies \citep[e.g., ][]{Mannucci2010a,yao2022a}. Solid lines represent the best fitted relations based on orthogonal distance regression (odr). Dashed lines represent the 1$\sigma$ distribution around the fitted FMR relations. Other markers are same to Fig. \ref{fig:msr_zoh}.}
   \label{fig:fmr_zoh}
 \end{figure*}

We also present the resolved fundamental metallicity relation \citep[FMR, e.g., ][]{Mannucci2010a,gao2018,gao2018a,Cresci2019,Curti2019} in Fig. \ref{fig:fmr_zoh}. The FMR, which relates the surface mass density ($\Sigma_*$), gas-phase metallicity ($Z$), and star formation rate surface density ($\Sigma_{\rm SFR}$), was developed to reduce the scatter in the mass-metallicity relation \citep[e.g.,][]{Tremonti2004,Mannucci2010a,Curti2019}. We calculate the combined quantity $\mu_{\alpha}$ as $\rm log(\Sigma_*) - \alpha log(\Sigma_{\rm SFR})$, where $\alpha = 0.32$, following the method in \cite{Mannucci2010a}. The solid and dashed lines in Fig. \ref{fig:fmr_zoh} represent the best-fit relations based on orthogonal distance regression (odr) and their 1$\sigma$ distribution, respectively.
{In the O3N2-based and N2S2-based FMRs, we observe that most of the star-forming knots at the interaction areas (marked as blue edgecolors) exhibited significantly lower gas-phase metallicity than expected for a given $\Sigma_*$ and $\Sigma_{\rm SFR}$, supporting the idea that the merger process can dilute the gas-phase metallicity by inflowing metal-poor gas.} However, the high gas-phase metallicity observed in other knots above the FMR may be explained by the metal enrichment caused by stellar activities. 

\subsection{Comparison with galaxy pairs}
\label{subsec:comp_pair}

\begin{figure}[t]
   \centering
   \includegraphics[width=0.45\textwidth]{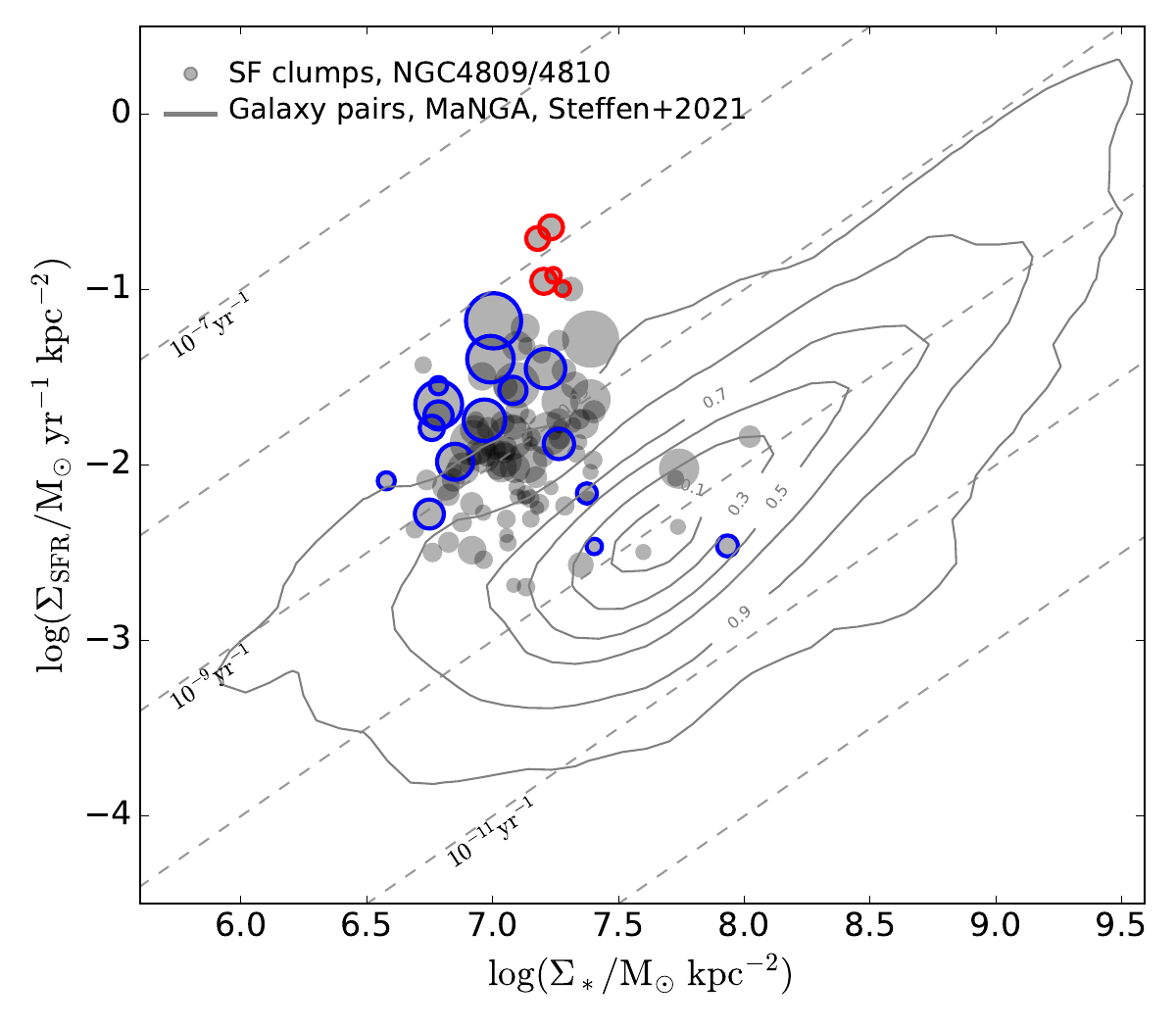}
   \caption{{Spatially resolved main sequence relation (rMSR) between the surface density of stellar mass ($\Sigma_*$) and the surface density of star formation rate ($\Sigma_{\rm SFR}$) for star-forming knots in NGC 4809/4810. The sizes of the markers correspond to the sizes of the knots. The gray contours indicate the distribution of $\Sigma_*$ and $\Sigma_{\rm SFR}$ of 54 star forming galaxy pairs selected by \cite{steffen2021} in MaNGA survey.} {We also provide the star formation knots around SN2011jm (with red edgecolors) and at the interaction area (with blue edgecolors), which are the same as those in Fig. \ref{fig:Ha}.} The dashed lines represent different specific star formation rate (sSFR, $\Sigma_{\rm SFR} / \Sigma_*$) levels. }
   \label{fig:msr_merger}
 \end{figure}

 {In previous studies, \cite{pan2019} and \cite{steffen2021} compared the specific SFR of star forming galaxy pairs and mass-matched control galaxies in the MaNGA survey. They found the star formation activities are triggered in the interaction processes, showing a $\sim$0.3 dex enhancement of sSFR within one effective radius. Here, we compare the star forming regions of NGC 4809/4810 with these galaxy pairs in Fig.\ref{fig:msr_merger}.   Gray contours indicate the stellar mass surface density and SFR surface density of 54 star forming galaxy pairs selected by \cite{pan2019} and \cite{steffen2021} from MaNGA survey. These galaxies are selected with log($M/M_{\odot}$) = 9.0 - 11.5 and log(sSFR/yr$^{-1}$) > -10.5. The stellar mass and emission line intensity at each spaxel are derived from the Pipe3D value added catalog\footnote{\url{https://www.sdss4.org/dr17/manga/manga-data/manga-pipe3d-value-added-catalog/}} \citep{sanchez2016,sanchez2016a,sanchez2018b}. The Pipe3D adopts a Salpeter IMF, while our analysis in Section \ref{subsec:muse_data} adopts the Chabrier IMF. So in order to convert stellar mass from a Salpeter IMF to Chabrier IMF, 0.2 dex has to be subtracted to the Pipe3D stellar mass. The significant star formation enhancement in galaxy pairs might be occurring at the coalescence phase (i.e., post merger) and the high stellar mass surface density region. We note the interaction areas between NGC 4809/4810 also show more than 1 dex higher SFR surface densities than the galaxy pairs at a similar stellar mass surface density. We suspect the interaction between the gas rich dwarf galaxies can trigger star formation activities at the less massive surface density region. Furthermore, the different spatial scale of the star formation clumps of NGC 4809/4810 and the spaxels in MaNGA galaxies might cause differences of sSFR.}

\subsection{Other properties of $\hii$ regions}
\label{subsec:hii_prop}

 \begin{figure*}[t]
   \centering
   \includegraphics[width=0.45\textwidth]{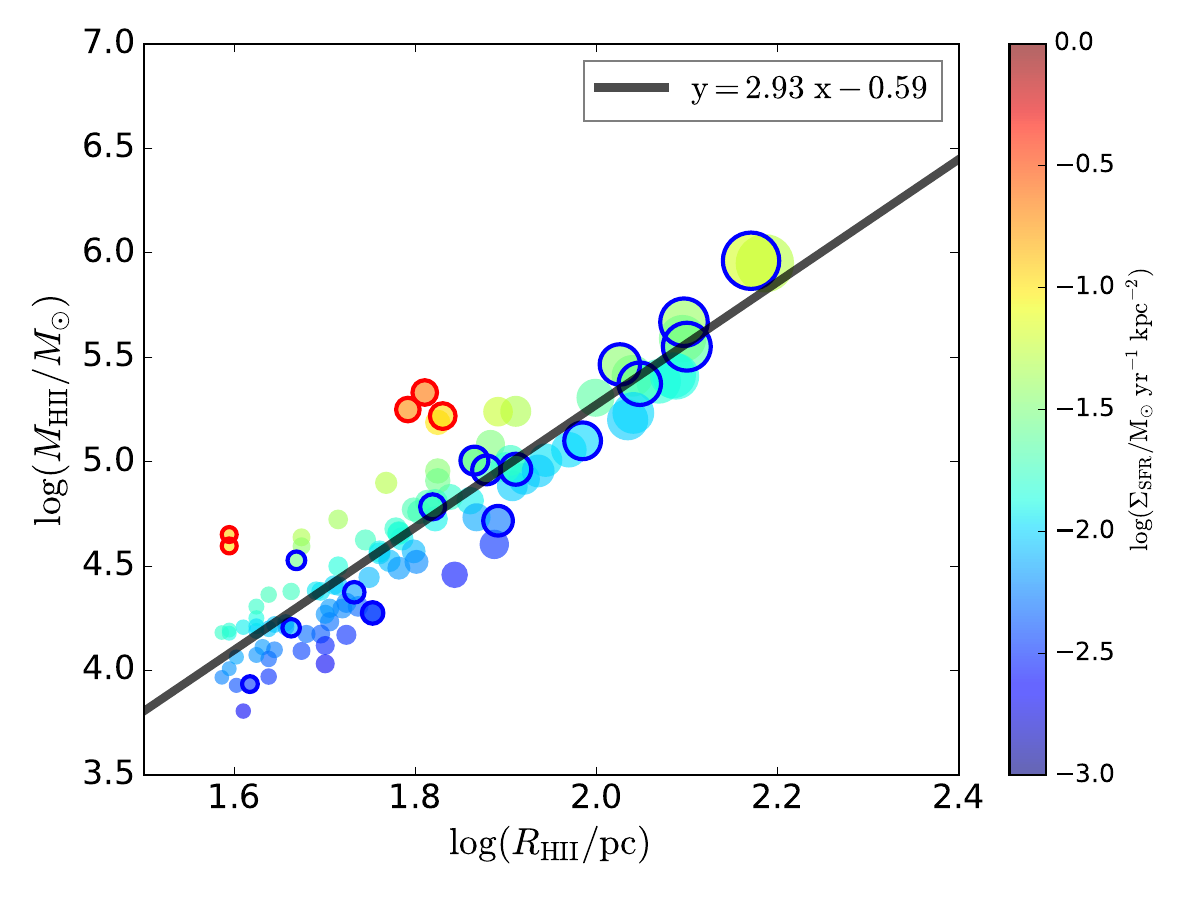}
   \includegraphics[width=0.45\textwidth]{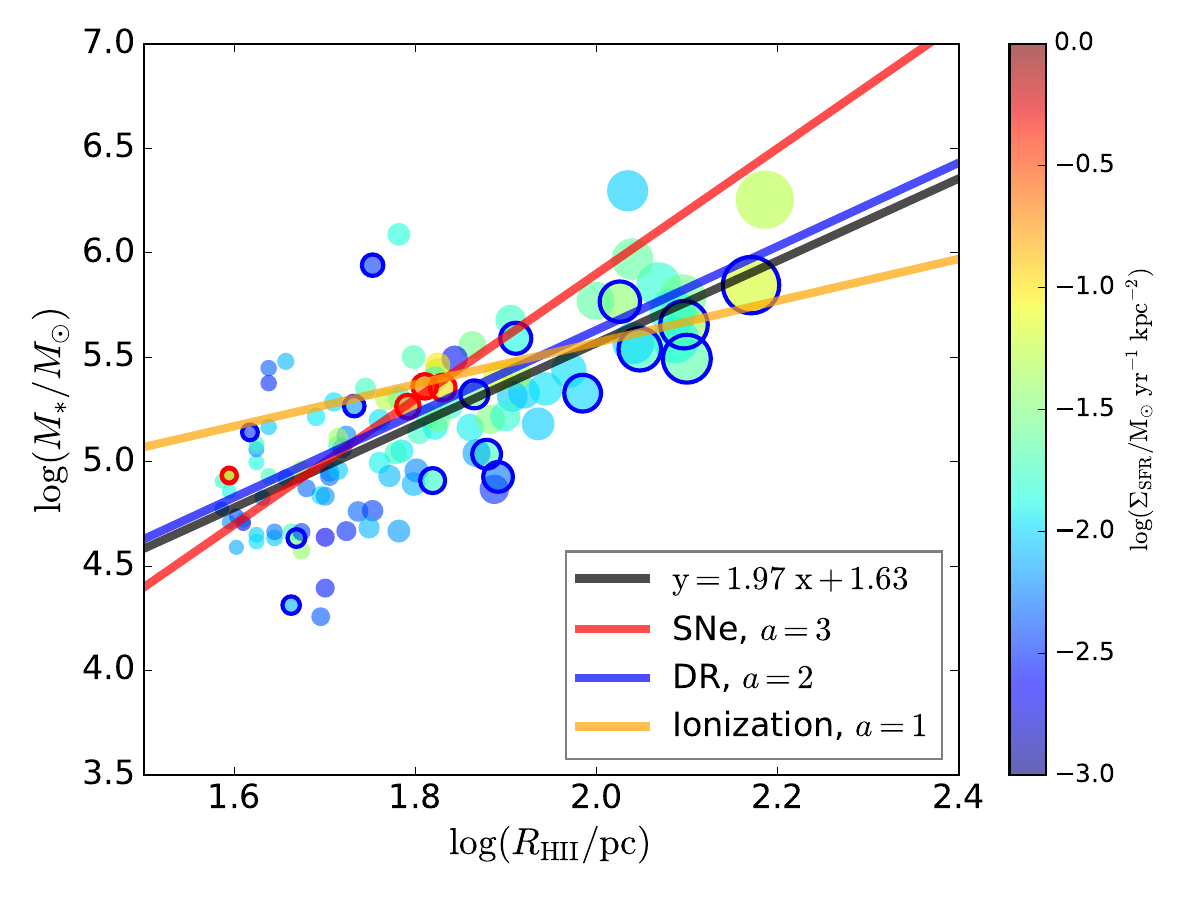}
   \caption{The relationship between the ionized gas mass and size ($M_{\hii} - R_{\hii}$) and between the stellar mass and size ($M_* - R_{\hii}$) for $\hii$ regions, color-coded by their SFR surface densities. The solid black lines represent the best-fit relations for $M_{\hii} - R_{\hii}$ and $M_* - R_{\hii}$, respectively. The solid colored lines indicate three different feedback mechanisms from \cite{Krumholz2018}: supernovae (SNe), direct radiation pressure (DR), and photoionization feedback (ionization), with their corresponding slopes ($a=3$, $a=2$, and $a=1$) displayed. Other markers are the same as in Fig. \ref{fig:msr_zoh}.}
   \label{fig:mass_size}
 \end{figure*}

\begin{figure*}[t]
   \centering
   \includegraphics[width=0.45\textwidth]{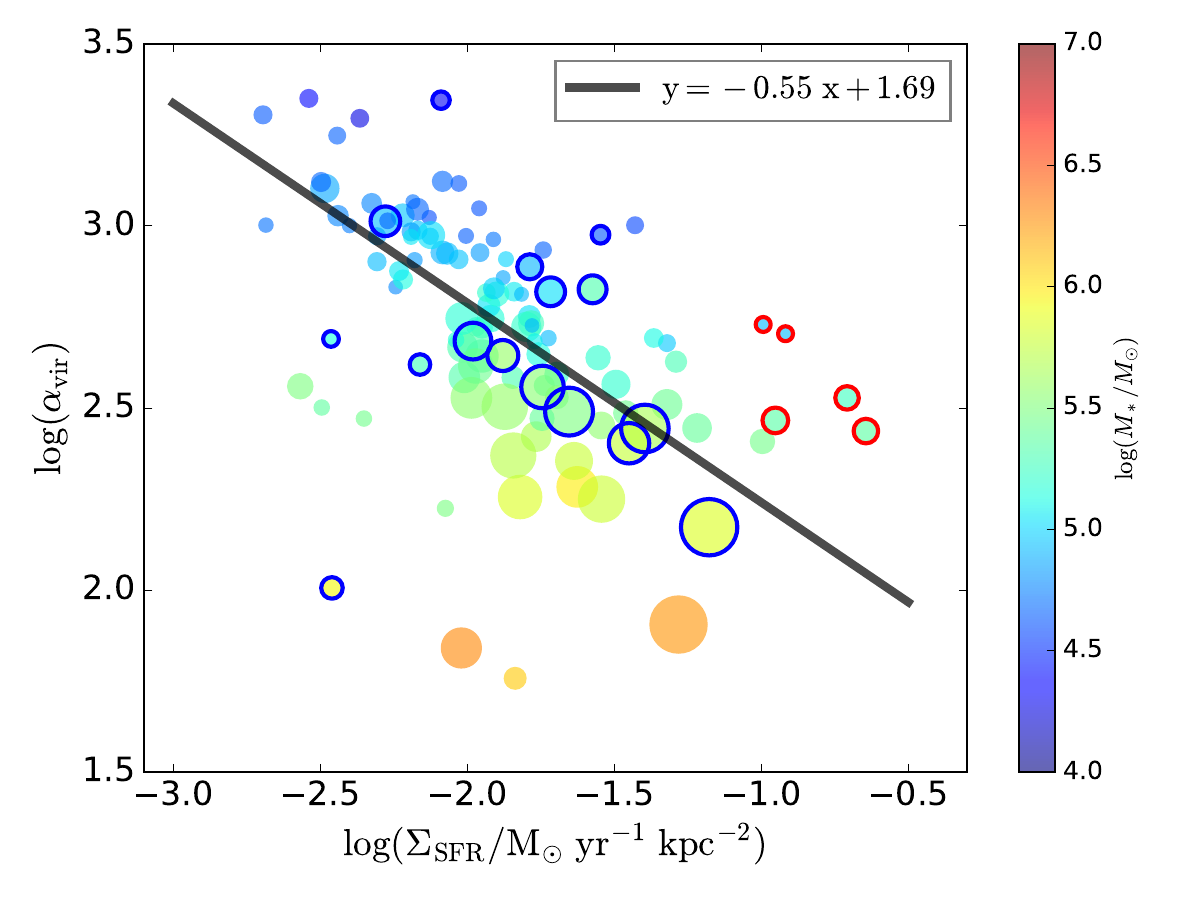}
   \includegraphics[width=0.45\textwidth]{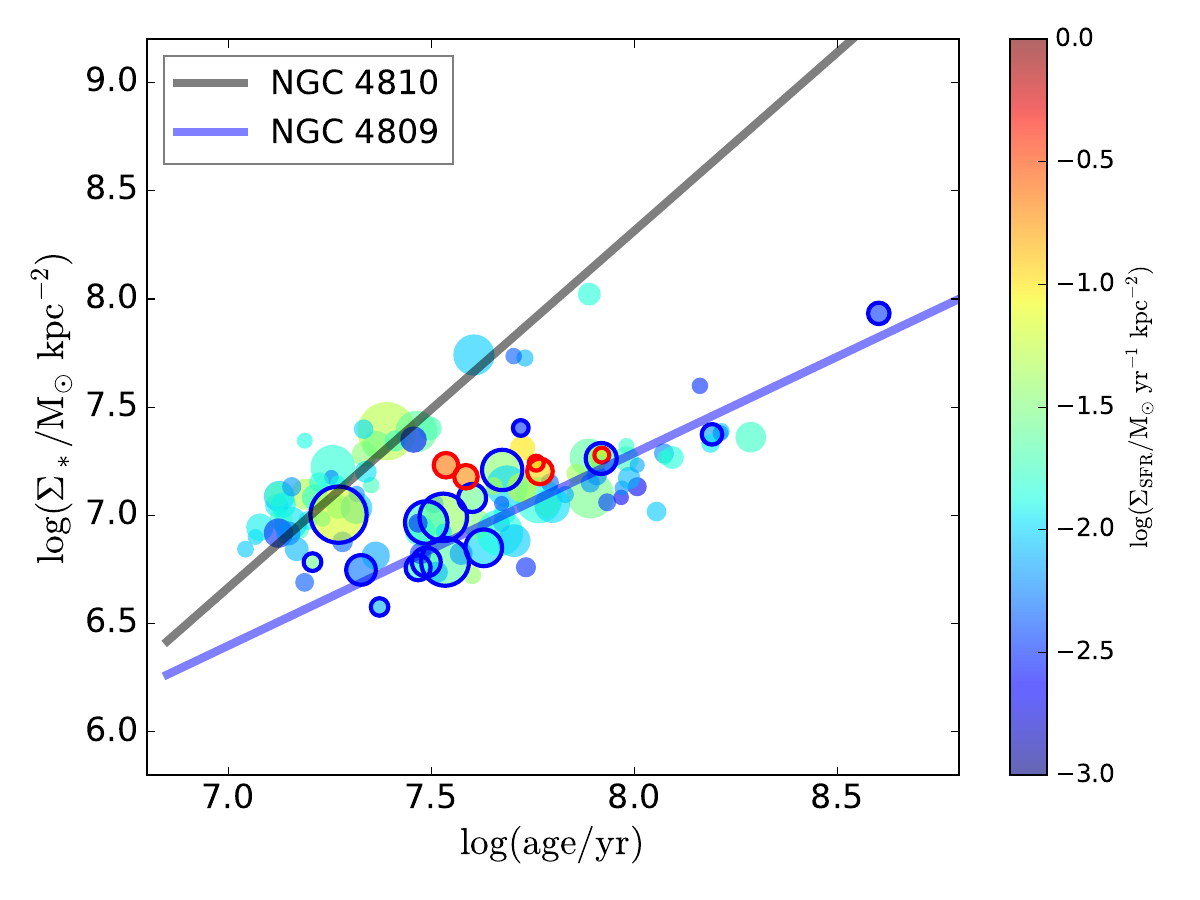}
   \caption{$Left$: the virial parameter versus the SFR surface density ($\alpha_{\rm vir} - \Sigma_{\rm SFR}$) of $\hii$ regions, colored by stellar mass. Solid line represents the best-fitted $\alpha_{\rm vir} - \Sigma_{\rm SFR}$ relation. $Right$: the stellar mass surface density - stellar age ($\Sigma_* - \rm age$) distribution of $\hii$ regions, colored by SFR surface densities. Solid lines mean the best-fitted relations of $\Sigma_* - \rm age$ in two component galaxies NGC 4809 and NGC 4810. Other markers are the same as in Fig. \ref{fig:msr_zoh}.}
   \label{fig:mass_age}
 \end{figure*}

 \begin{figure*}[t]
   \centering
   \includegraphics[width=0.45\textwidth]{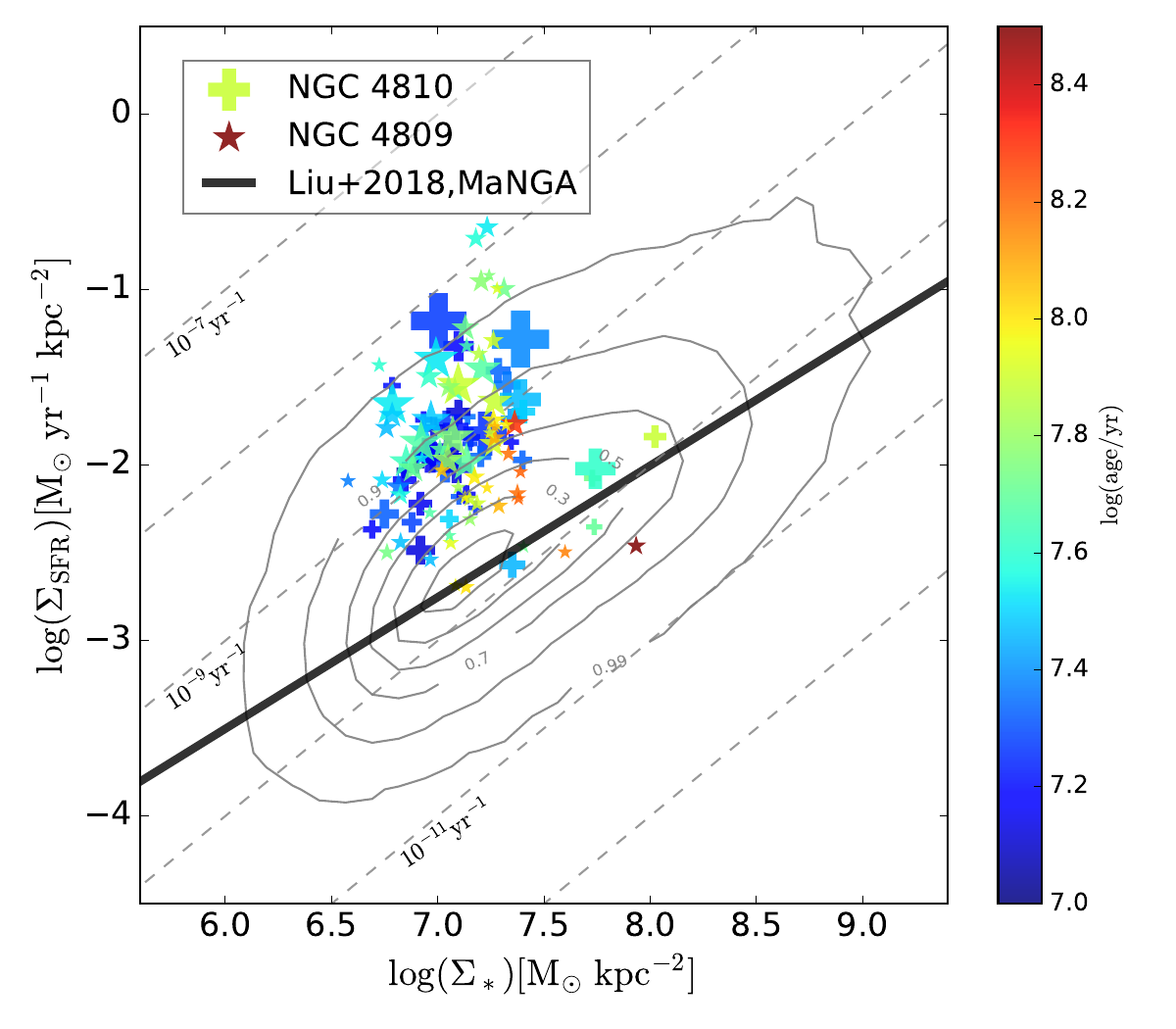}
   \includegraphics[width=0.45\textwidth]{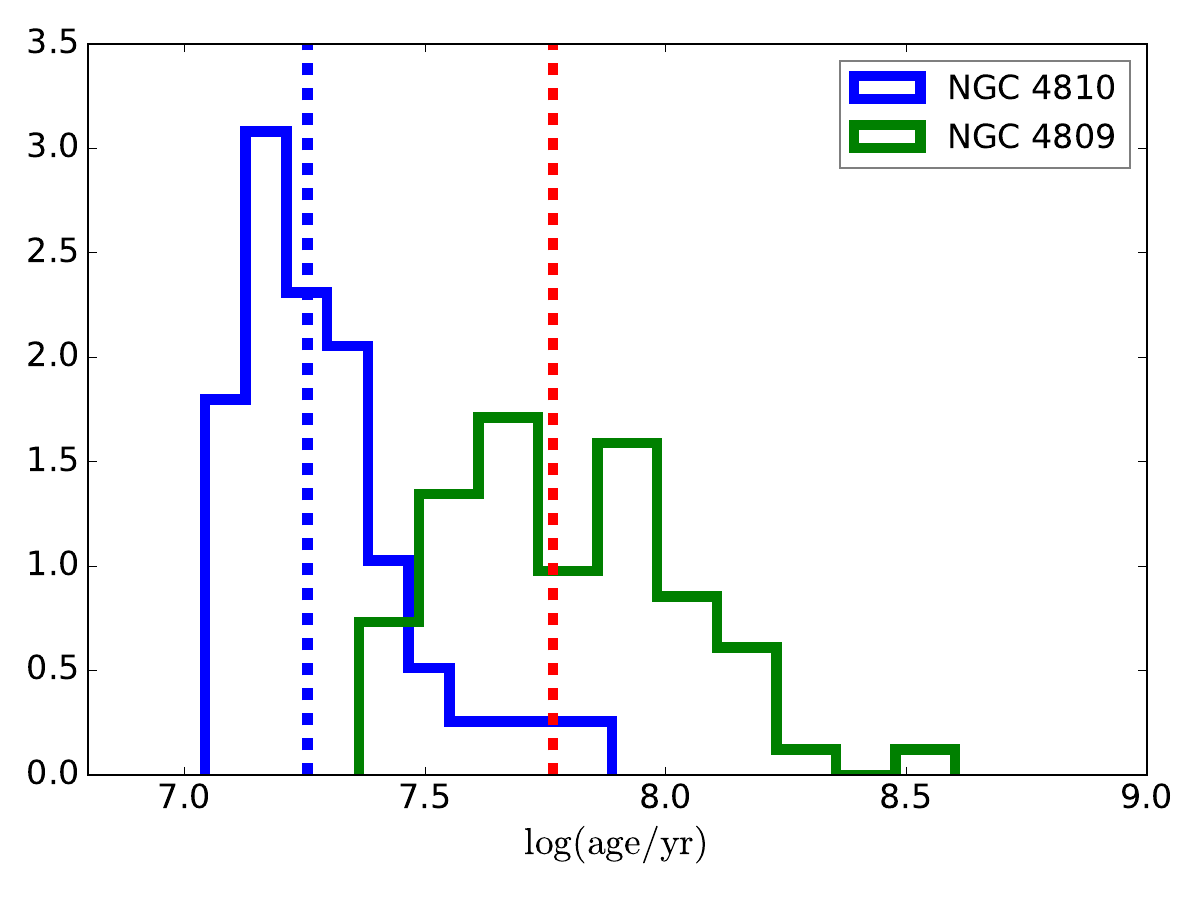}
   \caption{{$Left$: Stellar mass surface density - SFR surface density, $\Sigma_* - \Sigma_{\rm SFR}$ of star formation knots in NGC 4809 (marked as stars) and NGC 4810 (marked as pluses), colored by stellar age. The marker sizes are scaled to their knot sizes. $Right$: The histogram of stellar ages of star formation knots. Dashed lines represent their median values of $\rm log(age/yr)$, about 7.3 (NGC 4810) and 7.8 (NGC 4809), respectively.}}
   \label{fig:msr_age}
 \end{figure*}

From the attenuation-corrected $\ha$ luminosity, $L_{\ha}$, the radius of the $\hii$ region, $R_{\hii}$, and the velocity dispersion, $\sigma_{\rm v}$, we can determine several physical parameters, including the ionized gas mass of $\hii$ regions ($M_{\hii}$) and the virial parameter ($\alpha_{\rm vir}$). Using Eq.5 of \cite{zaragoza-cardiel2014}, we calculate the ionized gas mass, $M_{\hii}$, as follows:
\begin{equation}
M_{\hii} (\msun) = \frac{4}{3} \pi R_{\hii}^3 n_{\rm e} m_{\rm p} = 1.57 \times 10^{-17} (L_{\ha} R_{\hii}^3)^{0.5}.
\end{equation}
Here, $n_{\rm e}$ represents the electron density and $m_{\rm p} = 1.67 \times 10^{-27} \ \rm kg$ is the proton mass. The units for $L_{\ha}$ and $R_{\hii}$ are $\ergs$ and pc, respectively. We present the ionized gas mass -- size relation ($M_{\hii} - R_{\hii}$) of $\hii$ regions in the left panel of Fig.\ref{fig:mass_size}. The best-fitted relation is:
\begin{equation}
{\rm log}(M_{\hii}/\msun) = (2.93 \pm 0.12) \times {\rm log}(R_{\hii}/\rm pc) - (0.59 \pm 0.22),
\end{equation}
is shown as black solid line. 

We determine the power index $N$ of $M_{\hii} \propto R_{\hii}^N$ to be 2.93$\pm$0.12, which is approximately 3.0. This result suggests that the density of ionized gas remains relatively constant throughout most of the $\hii$ regions in NGC 4809/4810. However, we note that the regions around the supernova contain significantly more massive ($\sim 0.5$ dex) ionized gas for a given $\hii$ size, indicating higher densities of star-forming regions. This finding suggests that certain physical mechanisms, such as winds or shocks from stellar feedback, may compress the ionized gas within these regions.
Additionally, we plot the stellar mass--size relation ($M_* - R_{\hii}$) in the right panel of Fig.\ref{fig:mass_size}, and derive a best-fitted relation of
\begin{equation}
{\rm log}(M_*/\msun) = (1.97 \pm 0.18) \times {\rm log}(R_{\hii}/\rm pc) + (1.63 \pm 0.32).
\end{equation}
To determine the dominant feedback mechanism, we consider three different mechanisms from \cite{Krumholz2018}: supernovae (SNe, $a = 3$), direct radiation pressure (DR, $a = 2$), and photoionization feedback (ionization, $a = 1$), where $a$ is the slope of the relation ${\rm log}(M_/\msun) = a \times {\rm log}(R_{\hii}/\rm pc) + b$. Our analysis indicates that the best-fitted slope is $1.97 \pm 0.18$, consistent with the feedback mechanism of direct radiation pressure rather than SNe and photoionization. 
The average stellar mass surface density ($\Sigma_* = M_* / (\pi R)^2$) is about 13.6 $\msun \rm \ pc^{-2}$, smaller than the $\Sigma_{\rm DR} \sim 340 \msun \rm \ pc^{-2}$ derived by \cite{Krumholz2018}. Here, $\Sigma_{\rm DR}$ is the surface density below which direct radiation pressure becomes important. This result indicates that the direct radiation pressure rather than the photoionization or SNe dominates the ionized gas removal in these $\hii$ regions. 

We can use the velocity dispersion ($\sigma_{\rm v, \ha}$), ionized gas mass ($M_{\hii}$), and stellar mass in $\hii$ regions ($M_*$) to estimate the virial parameter of ionized gas, denoted by $\alpha_{\rm vir}$. Following \citep{bertoldi1992}, we can calculate $\alpha_{\rm vir} = \frac{5\sigma_{\rm v}^2 R_{\hii}}{G M}$, where $G \simeq 4.3 \times 10^{-3} \ {\rm pc} \ \msun^{-1} \ (\rm \kms)^2$ is the gravitational constant. We assume the mass in $\hii$ regions $M$ as the total mass of ionized gas and stellar mass, $M = M_{\hii} + M_*$. The virial parameter can be used to test whether the system is dominated by kinetic energy or gravitational potential \citep[e.g.,][]{Krumholz2005,Kauffmann2013}. If $\alpha_{\rm vir} >> 2$, the regions are dominated by internal pressure (e.g., ionized gas pressure, direct radiation pressure) instead of self-gravity. On the other hand, if $\alpha_{\rm vir} < 2$, the $\hii$ regions are dominated by gravity, meaning that the gas therein will be supercritical, unstable, and tend to collapse.

In Fig.\ref{fig:mass_age}, we plot the virial parameter versus the SFR surface density ($\alpha_{\rm vir} - \Sigma_{\rm SFR}$), with markers colored by stellar mass. We find that the virial parameters of ionized gas are much larger than 2, indicating that these $\hii$ regions are dominated by internal pressure and persistently expanding. Furthermore, the virial parameters are negatively correlated with the SFR surface densities, with a best-fitted relation of ${\rm log}(\alpha_{\rm vir}) = -0.55 \pm 0.07 \times {\rm log}(\Sigma_{\rm SFR}) + 1.69 \pm 0.13$. This negative correlation supports the scenario that regions with smaller virial parameters are efficient in forming new stars per unit area. We also observe that the virial parameter decreases as the mass of the $\hii$ regions increases. Conversely, the supernova-surrounding regions show much higher viral parameters for a given stellar mass, which might be caused by strong stellar feedback (direct radiation pressure) from the intense star formation activities therein.

In Fig.\ref{fig:mass_age}, we present the distribution of the stellar mass surface density - stellar age ($\Sigma_* - \rm age$) for $\hii$ regions in the galaxy merging system of NGC 4809 and NGC 4810. The solid lines represent the best-fitted relations of $\Sigma_* - \rm age$ in both component galaxies. The relation for NGC 4809 is given by the equation:
\begin{equation}
{\rm log}(\Sigma_*) = (1.03 \pm 0.11) \times {\rm log(age)} - (1.05 \pm 0.84),
\end{equation}
whereas the relation for NGC 4810 is given by:
\begin{equation}
{\rm log}(\Sigma_*) = (1.87 \pm 0.20) \times {\rm log(age)} - (6.52 \pm 1.43).
\end{equation}
Here, $\Sigma_*$ and age are given in units of $\msun \rm \ kpc^{-2}$ and years, respectively. The power index of age in NGC 4809 is 1.03, which is close to 1.0, indicating that the stellar mass is expected to increase with the stellar age, i.e., $M_* \propto \rm age$, at a fixed area size. This suggests a scenario in which the SFR is constant with time \citep{Krumholz2018}. However, the power index in NGC 4810 is 1.87, which is much larger than 1.0 and inconsistent with a constant star formation history. This difference in power index indicates that the stellar mass assembly process in NGC 4810 is more efficient than in NGC 4809. 
Most of the star formation knots show a young stellar age, i.e., younger than 100 Myr, indicating that the galaxy merging system has created stars in the past 100 Myr. This is consistent with the time at which star formation is triggered by the interaction of galaxies. Additionally, the regions at the interaction area and the supernova-surrounding regions show relatively higher stellar mass surface densities at a given stellar age, confirming their enhanced star formation activities.

{To compare the efficiency of stellar mass assembly in different systems, we present the relation between stellar mass surface density and SFR surface density colored by their stellar ages, as shown in $left$ panel of Fig. \ref{fig:msr_age}. We also present the histogram of stellar ages of these star formation knots in the $right$ panel of Fig. \ref{fig:msr_age}. It is important to note that the stellar ages in NGC 4810 are consistently smaller than those in NGC 4809, with median $\rm log(age/yr)$ values of 7.3 and 7.8, respectively. We suggest a possible scenario in which the interaction first destabilized the molecular gas cloud in NGC 4809, triggering star formation and then enriching the metallicity of ISM by stellar activities (e.g., supernovae, winds) in the past 100 Myr to 30 Myr. Subsequently, the interaction likely transported significant amounts of metal-enriched molecular and ionized gas into NGC 4810, thereby stimulating the more efficient creation of new stars at a shorter timescale of 30 Myr to 10 Myr. Then these stellar activities create and eject metal into the surrounding ISM. The metal-richer ISM in NGC 4810 (shown at the N2S2-based metallicity map in Fig.\ref{fig:zoh}), and the more efficient stellar mass assembly process in NGC 4810 (see $right$ panel of Fig. \ref{fig:mass_age}) support this scenario. However, future observations of molecular gas will be necessary to verify this scenario. For example, the molecular gas mass fraction and its spatial distribution can provide the gas density and star formation efficiency ($\rm M_{\rm gas}/SFR$, SFE) in these two galaxy components, or even at/around different knots at high spatial resolution scales. Then we can check that the average SFE in NGC 4810 is whether higher than in NGC 4809 or not. The kinematics and dispersion of molecular gas will also shed light on these effects, such as  transportation, compression, and disruption, of interaction/merging processes on molecular clouds.} 


\section{Summary}
\label{sec:sum}

This study investigates the ionized gas properties and star formation activities in the merging dwarf galaxy system NGC 4809/4810, using the VLT/MUSE IFU data. Our main results are as follows:

\begin{itemize}
\item We identify 112 $\ha$ emission knots within the NGC 4809/4810 system, in which the gas ionization is primarily caused by star formation, with a few highly-ionized regions likely resulting from high-energy photons emitted by young and massive stars/clusters.

\item The distribution of gas-phase metallicity is inhomogeneous across the two galaxies, with a mixture of metal-poor and metal-rich ionized gas. Star-forming knots at the interaction area show deficient metallicity based on the O3N2 and N2S2 diagnostics, which could be due to the dilution by metal-poor gas inflows during the merging process of two dwarf galaxies.

\item When comparing the SFR of NGC 4809/4810 with that of normal star-forming galaxies and the resolved stellar mass-SFR relation, we find that the interaction area and the ring around the supernova contained the highest SFRs and SFR/rMSR tatios, indicating that the merging process of two dwarf galaxies can trigger star formation activities at the interaction area, even in a metal-poor environment.

\item The slope of the ionized gas mass-size relation of $\hii$ regions is 2.93 $\pm$ 0.12, indicating that the ionized gas density is nearly constant. However, the regions around the supernova contain much denser ionized gas. The slope of the stellar mass-size relation is 1.97 $\pm$ 0.18, suggesting that direct radiation pressure, rather than photoionization or supernovae, dominates ionized gas removal in these $\hii$ regions.

\item The high virial parameters ($>>2$) of the ionized gas indicate that these $\hii$ regions are dominated by internal pressure and are persistently expanding. The negative correlation between virial parameters and SFR surface densities supports the scenario that regions with smaller virial parameters are efficient at forming new stars per unit area.

\item We detect two different relations between stellar mass surface density and stellar age in NGC 4809 and NGC 4810. In NGC 4809, the star formation rate remains constant with time, whereas in NGC 4810, the stellar mass assembly process is more efficient. These might indicate that galaxy interaction first destabilize the molecular gas cloud in NGC 4809, then transport molecular gas into NGC 4810 and create new stars.
\end{itemize}

In conclusion, the merging of two dwarf galaxies can induce starburst activities in the interaction areas, despite their location in metal-deficient environments. Our findings shed light on the mechanisms driving starbursts in dwarf galaxies. {We recommend  performing the sub-millimeter observations (e.g., NOEMA, ALMA), such as the cold diffuse molecular gas tracer \citep[e.g., CO(1-0), CO(2-1), ][]{gao1999,Kennicutt2012} and the warm denser gas tracer \citep[e.g., CO(6-5), HCN(1-0), HCN(4-3), ][]{Gao2004,lu2014,zhang2014,cao2018}, to investigate the transportation, compression, and disruption of gas clouds, as well as the star formation efficiency in this merging galaxy system. Furthermore, with the successful launch of the JWST, it will be possible to probe the properties of the small dust grains and polycyclic aromatic hydrocarbons (PAHs) at a scale of $<$ 20 pc (0.25$\arcsec$). This will reveal bubbles/shell-like structures at/around young stars/clusters and interaction areas to study the effect of stellar feedback on the gas consumption, dust creation, and metal enrichment.}

\begin{acknowledgements}
We thank the referee very much for careful reading and valuable suggestions.
Y.L.G acknowledge the grant from the National Natural Science Foundation of China (No. 12103023).
This work is supported by the National Natural Science Foundation of China (No. 12192222, 12192220 and 12121003). 
We acknowledge the science research grants from the China Manned Space Project with NO. CMS-CSST-2021-A05.
G.L acknowledges the support from the China Manned Space Project (No. CMS-CSST-2021-A06, CMSCSST-2021-A07, and the 2nd-stage CSST science project: {\em Investigation of small-scale structures in galaxies and forecasting of observations}), the National Natural Science Foundation of China (No. 12273036, 11421303), the Fundamental Research Funds for the Central Universities (No. WK3440000005), and the lateral fund from Shanghai Astronomical Observatory (No. EF2030220007).
This project makes use of the MaNGA-Pipe3D dataproducts. We thank the IA-UNAM MaNGA team for creating this catalogue, and the Conacyt Project CB-285080 for supporting them. 

\end{acknowledgements}

%
%
\bibliographystyle{aa}
\bibliography{ngc4809}

\begin{thebibliography}{91}
\expandafter\ifx\csname natexlab\endcsname\relax\def\natexlab#1{#1}\fi

\bibitem[{{Allende Prieto} {et~al.}(2001){Allende Prieto}, {Lambert}, \&
  {Asplund}}]{Allende2001}
{Allende Prieto}, C., {Lambert}, D.~L., \& {Asplund}, M. 2001, \apjl, 556, L63

\bibitem[{Alloin {et~al.}(1979)Alloin, Collin-Souffrin, Joly, \&
  Vigroux}]{Alloin1979}
Alloin, D., Collin-Souffrin, S., Joly, M., \& Vigroux, L. 1979, \aap, 78, 200

\bibitem[{{Bacon} {et~al.}(2010){Bacon}, {Accardo}, {Adjali}, {Anwand},
  {Bauer}, {Biswas}, {Blaizot}, {Boudon}, {Brau-Nogue}, {Brinchmann},
  {Caillier}, {Capoani}, {Carollo}, {Contini}, {Couderc}, {Daguis{\'e}},
  {Deiries}, {Delabre}, {Dreizler}, {Dubois}, {Dupieux}, {Dupuy}, {Emsellem},
  {Fechner}, {Fleischmann}, {Fran{\c{c}}ois}, {Gallou}, {Gharsa}, {Glindemann},
  {Gojak}, {Guiderdoni}, {Hansali}, {Hahn}, {Jarno}, {Kelz}, {Koehler},
  {Kosmalski}, {Laurent}, {Le Floch}, {Lilly}, {Lizon}, {Loupias}, {Manescau},
  {Monstein}, {Nicklas}, {Olaya}, {Pares}, {Pasquini}, {P{\'e}contal-Rousset},
  {Pell{\'o}}, {Petit}, {Popow}, {Reiss}, {Remillieux}, {Renault}, {Roth},
  {Rupprecht}, {Serre}, {Schaye}, {Soucail}, {Steinmetz}, {Streicher}, {Stuik},
  {Valentin}, {Vernet}, {Weilbacher}, {Wisotzki}, \& {Yerle}}]{Bacon2010}
{Bacon}, R., {Accardo}, M., {Adjali}, L., {et~al.} 2010, in Society of
  Photo-Optical Instrumentation Engineers (SPIE) Conference Series, Vol. 7735,
  Ground-based and Airborne Instrumentation for Astronomy III, ed. I.~S.
  {McLean}, S.~K. {Ramsay}, \& H.~{Takami}, 773508

\bibitem[{{Baldwin} {et~al.}(1981){Baldwin}, {Phillips}, \&
  {Terlevich}}]{Baldwin1981}
{Baldwin}, A., {Phillips}, M.~M., \& {Terlevich}, R. 1981, \pasp, 93, 817

\bibitem[{{Belfiore} {et~al.}(2016){Belfiore}, {Maiolino}, {Maraston},
  {Emsellem}, {Bershady}, {Masters}, {Yan}, {Bizyaev}, {Boquien}, {Brownstein},
  {Bundy}, {Drory}, {Heckman}, {Law}, {Roman-Lopes}, {Pan}, {Stanghellini},
  {Thomas}, {Weijmans}, \& {Westfall}}]{Belfiore2016a}
{Belfiore}, F., {Maiolino}, R., {Maraston}, C., {et~al.} 2016, \mnras, 461,
  3111

\bibitem[{{Bell} {et~al.}(2003){Bell}, {McIntosh}, {Katz}, \&
  {Weinberg}}]{Bell2003}
{Bell}, E.~F., {McIntosh}, D.~H., {Katz}, N., \& {Weinberg}, M.~D. 2003, \apjs,
  149, 289

\bibitem[{{Bertoldi} \& {McKee}(1992)}]{bertoldi1992}
{Bertoldi}, F. \& {McKee}, C.~F. 1992, \apj, 395, 140

\bibitem[{{Bresolin}(2019)}]{bresolin2019}
{Bresolin}, F. 2019, \mnras, 488, 3826

\bibitem[{Brinchmann {et~al.}(2004)Brinchmann, Charlot, White, Tremonti,
  Kauffmann, Heckman, \& Brinkmann}]{Brinchmann2004a}
Brinchmann, J., Charlot, S., White, S. D.~M., {et~al.} 2004, \mnras, 351, 1151

\bibitem[{{Bruzual} \& {Charlot}(2003)}]{Bruzual2003}
{Bruzual}, G. \& {Charlot}, S. 2003, \mnras, 344, 1000

\bibitem[{{Cair{\'o}s} {et~al.}(2009){Cair{\'o}s}, {Caon}, {Papaderos},
  {Kehrig}, {Weilbacher}, {Roth}, \& {Zurita}}]{cairos2009}
{Cair{\'o}s}, L.~M., {Caon}, N., {Papaderos}, P., {et~al.} 2009, \apj, 707,
  1676

\bibitem[{{Calzetti} {et~al.}(2000){Calzetti}, {Armus}, {Bohlin}, {Kinney},
  {Koornneef}, \& {Storchi-Bergmann}}]{Calzetti2000a}
{Calzetti}, D., {Armus}, L., {Bohlin}, R.~C., {et~al.} 2000, \apj, 533, 682

\bibitem[{{Cao} {et~al.}(2018){Cao}, {Lu}, {Xu}, {Zhao}, {Kalari}, {Gao},
  {Charmandaris}, {Diaz Santos}, {van der Werf}, {Cao}, {Wu}, {Inami}, \&
  {Evans}}]{cao2018}
{Cao}, T., {Lu}, N., {Xu}, C.~K., {et~al.} 2018, \apj, 866, 117

\bibitem[{{Casasola} {et~al.}(2004){Casasola}, {Bettoni}, \&
  {Galletta}}]{casasola2004a}
{Casasola}, V., {Bettoni}, D., \& {Galletta}, G. 2004, \aap, 422, 941

\bibitem[{{Chabrier}(2003)}]{Chabrier2003}
{Chabrier}, G. 2003, \pasp, 115, 763

\bibitem[{{Cheung} {et~al.}(2016){Cheung}, {Bundy}, {Cappellari}, {Peirani},
  {Rujopakarn}, {Westfall}, {Yan}, {Bershady}, {Greene}, {Heckman}, {Drory},
  {Law}, {Masters}, {Thomas}, {Wake}, {Weijmans}, {Rubin}, {Belfiore},
  {Vulcani}, {Chen}, {Zhang}, {Gelfand}, {Bizyaev}, {Roman-Lopes}, \&
  {Schneider}}]{Cheung2016May}
{Cheung}, E., {Bundy}, K., {Cappellari}, M., {et~al.} 2016, \nat, 533, 504

\bibitem[{{Cid Fernandes} {et~al.}(2005){Cid Fernandes}, {Mateus}, {Sodr{\'e}},
  {Stasi{\'n}ska}, \& {Gomes}}]{CidFernandes2005}
{Cid Fernandes}, R., {Mateus}, A., {Sodr{\'e}}, L., {Stasi{\'n}ska}, G., \&
  {Gomes}, J.~M. 2005, \mnras, 358, 363

\bibitem[{{Cid Fernandes} {et~al.}(2010){Cid Fernandes}, {Stasi{\'n}ska},
  {Schlickmann}, {Mateus}, {Vale Asari}, {Schoenell}, \&
  {Sodr{\'e}}}]{cidfernandes2010}
{Cid Fernandes}, R., {Stasi{\'n}ska}, G., {Schlickmann}, M.~S., {et~al.} 2010,
  \mnras, 403, 1036

\bibitem[{{Cresci} {et~al.}(2019){Cresci}, {Mannucci}, \& {Curti}}]{Cresci2019}
{Cresci}, G., {Mannucci}, F., \& {Curti}, M. 2019, \aap, 627, A42

\bibitem[{Curti {et~al.}(2019)Curti, Mannucci, Cresci, \& Maiolino}]{Curti2019}
Curti, M., Mannucci, F., Cresci, G., \& Maiolino, R. 2019
  [\eprint{1910.00597v1}]

\bibitem[{{Daddi} {et~al.}(2007){Daddi}, {Dickinson}, {Morrison}, {Chary},
  {Cimatti}, {Elbaz}, {Frayer}, {Renzini}, {Pope}, {Alexander}, {Bauer},
  {Giavalisco}, {Huynh}, {Kurk}, \& {Mignoli}}]{Daddi2007}
{Daddi}, E., {Dickinson}, M., {Morrison}, G., {et~al.} 2007, \apj, 670, 156

\bibitem[{{Dey} {et~al.}(2019){Dey}, {Schlegel}, {Lang}, {Blum}, {Burleigh},
  {Fan}, {Findlay}, {Finkbeiner}, {Herrera}, {Juneau}, {Landriau}, {Levi},
  {McGreer}, {Meisner}, {Myers}, {Moustakas}, {Nugent}, {Patej}, {Schlafly},
  {Walker}, {Valdes}, {Weaver}, {Y{\`e}che}, {Zou}, {Zhou}, {Abareshi},
  {Abbott}, {Abolfathi}, {Aguilera}, {Alam}, {Allen}, {Alvarez}, {Annis},
  {Ansarinejad}, {Aubert}, {Beechert}, {Bell}, {BenZvi}, {Beutler}, {Bielby},
  {Bolton}, {Brice{\~n}o}, {Buckley-Geer}, {Butler}, {Calamida}, {Carlberg},
  {Carter}, {Casas}, {Castander}, {Choi}, {Comparat}, {Cukanovaite}, {Delubac},
  {DeVries}, {Dey}, {Dhungana}, {Dickinson}, {Ding}, {Donaldson}, {Duan},
  {Duckworth}, {Eftekharzadeh}, {Eisenstein}, {Etourneau}, {Fagrelius},
  {Farihi}, {Fitzpatrick}, {Font-Ribera}, {Fulmer}, {G{\"a}nsicke},
  {Gaztanaga}, {George}, {Gerdes}, {Gontcho}, {Gorgoni}, {Green}, {Guy},
  {Harmer}, {Hernandez}, {Honscheid}, {Huang}, {James}, {Jannuzi}, {Jiang},
  {Joyce}, {Karcher}, {Karkar}, {Kehoe}, {Kneib}, {Kueter-Young}, {Lan},
  {Lauer}, {Le Guillou}, {Le Van Suu}, {Lee}, {Lesser}, {Perreault Levasseur},
  {Li}, {Mann}, {Marshall}, {Mart{\'\i}nez-V{\'a}zquez}, {Martini}, {du Mas des
  Bourboux}, {McManus}, {Meier}, {M{\'e}nard}, {Metcalfe},
  {Mu{\~n}oz-Guti{\'e}rrez}, {Najita}, {Napier}, {Narayan}, {Newman}, {Nie},
  {Nord}, {Norman}, {Olsen}, {Paat}, {Palanque-Delabrouille}, {Peng},
  {Poppett}, {Poremba}, {Prakash}, {Rabinowitz}, {Raichoor}, {Rezaie},
  {Robertson}, {Roe}, {Ross}, {Ross}, {Rudnick}, {Safonova}, {Saha},
  {S{\'a}nchez}, {Savary}, {Schweiker}, {Scott}, {Seo}, {Shan}, {Silva},
  {Slepian}, {Soto}, {Sprayberry}, {Staten}, {Stillman}, {Stupak}, {Summers},
  {Sien Tie}, {Tirado}, {Vargas-Maga{\~n}a}, {Vivas}, {Wechsler}, {Williams},
  {Yang}, {Yang}, {Yapici}, {Zaritsky}, {Zenteno}, {Zhang}, {Zhang}, {Zhou}, \&
  {Zhou}}]{dey2019}
{Dey}, A., {Schlegel}, D.~J., {Lang}, D., {et~al.} 2019, \aj, 157, 168

\bibitem[{{Dopita} {et~al.}(2016){Dopita}, {Kewley}, {Sutherland}, \&
  {Nicholls}}]{dopita2016}
{Dopita}, M.~A., {Kewley}, L.~J., {Sutherland}, R.~S., \& {Nicholls}, D.~C.
  2016, \apss, 361, 61

\bibitem[{{Elbaz} {et~al.}(2018){Elbaz}, {Leiton}, {Nagar}, {Okumura},
  {Franco}, {Schreiber}, {Pannella}, {Wang}, {Dickinson}, {D{\'\i}az-Santos},
  {Ciesla}, {Daddi}, {Bournaud}, {Magdis}, {Zhou}, \& {Rujopakarn}}]{elbaz2018}
{Elbaz}, D., {Leiton}, R., {Nagar}, N., {et~al.} 2018, \aap, 616, A110

\bibitem[{{Espada} {et~al.}(2018){Espada}, {Martin}, {Verley}, {Pettitt},
  {Matsushita}, {Argudo-Fern{\'a}ndez}, {Randriamanakoto}, {Hsieh}, {Saito},
  {Miura}, {Kawana}, {Sabater}, {Verdes-Montenegro}, {Ho}, {Kawabe}, \&
  {Iono}}]{Espada2018}
{Espada}, D., {Martin}, S., {Verley}, S., {et~al.} 2018, \apj, 866, 77

\bibitem[{{Fabian}(2012)}]{fabian2012}
{Fabian}, A.~C. 2012, \araa, 50, 455

\bibitem[{{Fang} {et~al.}(2012){Fang}, {Faber}, {Salim}, {Graves}, \&
  {Rich}}]{Fang2012}
{Fang}, J.~J., {Faber}, S.~M., {Salim}, S., {Graves}, G.~J., \& {Rich}, R.~M.
  2012, \apj, 761, 23

\bibitem[{{Gao} {et~al.}(2018{\natexlab{a}}){Gao}, {Bao}, {Yuan}, {Kong},
  {Zou}, {Zhou}, {Gu}, {Lin}, {Liang}, \& {Huang}}]{gao2018}
{Gao}, Y., {Bao}, M., {Yuan}, Q., {et~al.} 2018{\natexlab{a}}, \apj, 869, 15

\bibitem[{{Gao} {et~al.}(2022){Gao}, {Gu}, {Shi}, {Zhou}, {Bao}, {Yu}, {Zhang},
  {Wang}, {Madden}, {Hayes}, {Lu}, \& {Xu}}]{gao2022a}
{Gao}, Y., {Gu}, Q., {Shi}, Y., {et~al.} 2022, \aap, 661, A136

\bibitem[{{Gao} \& {Solomon}(1999)}]{gao1999}
{Gao}, Y. \& {Solomon}, P.~M. 1999, \apjl, 512, L99

\bibitem[{{Gao} \& {Solomon}(2004)}]{Gao2004}
{Gao}, Y. \& {Solomon}, P.~M. 2004, \apj, 606, 271

\bibitem[{{Gao} {et~al.}(2018{\natexlab{b}}){Gao}, {Wang}, {Kong}, {Lin},
  {Liu}, {Liu}, {Liu}, {Hu}, {Berhane Teklu}, {Chen}, \& {Zhao}}]{gao2018a}
{Gao}, Y., {Wang}, E., {Kong}, X., {et~al.} 2018{\natexlab{b}}, \apj, 868, 89

\bibitem[{{Garc{\'\i}a-Benito} \&
  {P{\'e}rez-Montero}(2012)}]{garcia-benito2012}
{Garc{\'\i}a-Benito}, R. \& {P{\'e}rez-Montero}, E. 2012, \mnras, 423, 406

\bibitem[{{Goodman} {et~al.}(2009){Goodman}, {Rosolowsky}, {Borkin}, {Foster},
  {Halle}, {Kauffmann}, \& {Pineda}}]{Goodman2009Jan}
{Goodman}, A.~A., {Rosolowsky}, E.~W., {Borkin}, M.~A., {et~al.} 2009, \nat,
  457, 63

\bibitem[{{Harrison} {et~al.}(2018){Harrison}, {Costa}, {Tadhunter},
  {Fl{\"u}tsch}, {Kakkad}, {Perna}, \& {Vietri}}]{harrison2018}
{Harrison}, C.~M., {Costa}, T., {Tadhunter}, C.~N., {et~al.} 2018, Nature
  Astronomy, 2, 198

\bibitem[{{Hayes} {et~al.}(2013){Hayes}, {{\"O}stlin}, {Schaerer}, {Verhamme},
  {Mas-Hesse}, {Adamo}, {Atek}, {Cannon}, {Duval}, {Guaita}, {Herenz}, {Kunth},
  {Laursen}, {Melinder}, {Orlitov{\'a}}, {Ot{\'\i}-Floranes}, \&
  {Sandberg}}]{hayes2013}
{Hayes}, M., {{\"O}stlin}, G., {Schaerer}, D., {et~al.} 2013, \apjl, 765, L27

\bibitem[{{Heckman} {et~al.}(2005){Heckman}, {Hoopes}, {Seibert}, {Martin},
  {Salim}, {Rich}, {Kauffmann}, {Charlot}, {Barlow}, {Bianchi}, {Byun},
  {Donas}, {Forster}, {Friedman}, {Jelinsky}, {Lee}, {Madore}, {Malina},
  {Milliard}, {Morrissey}, {Neff}, {Schiminovich}, {Siegmund}, {Small},
  {Szalay}, {Welsh}, \& {Wyder}}]{heckman2005}
{Heckman}, T.~M., {Hoopes}, C.~G., {Seibert}, M., {et~al.} 2005, \apjl, 619,
  L35

\bibitem[{{Heckman} {et~al.}(2001){Heckman}, {Sembach}, {Meurer}, {Leitherer},
  {Calzetti}, \& {Martin}}]{heckman2001}
{Heckman}, T.~M., {Sembach}, K.~R., {Meurer}, G.~R., {et~al.} 2001, \apj, 558,
  56

\bibitem[{{Hopkins} \& {Quataert}(2010)}]{Hopkins2009}
{Hopkins}, P.~F. \& {Quataert}, E. 2010, \mnras, 407, 1529

\bibitem[{Howerton {et~al.}(2011)Howerton, Drake, Djorgovski, Mahabal, Graham,
  Williams, Prieto, Catelan, McNaught, Garradd, Beshore, Larson, Christensen,
  Brimacombe, Luppi, Buzzi, Baroni, Concari, Foglia, Galli, Tombelli, Foley, \&
  Fong}]{howerton2011a}
Howerton, S., Drake, A.~J., Djorgovski, S.~G., {et~al.} 2011, Central Bureau
  Electronic Telegrams, 2962, 1

\bibitem[{{Israel} {et~al.}(2015){Israel}, {Rosenberg}, \& {van der
  Werf}}]{Israel2015}
{Israel}, F.~P., {Rosenberg}, M.~J.~F., \& {van der Werf}, P. 2015, \aap, 578,
  A95

\bibitem[{{Izotov} {et~al.}(2006){Izotov}, {Stasi{\'n}ska}, {Meynet}, {Guseva},
  \& {Thuan}}]{Izotov2006}
{Izotov}, Y.~I., {Stasi{\'n}ska}, G., {Meynet}, G., {Guseva}, N.~G., \&
  {Thuan}, T.~X. 2006, \aap, 448, 955

\bibitem[{{James} {et~al.}(2020){James}, {Kumari}, {Emerick}, {Koposov},
  {McQuinn}, {Stark}, {Belokurov}, \& {Maiolino}}]{james2020}
{James}, B.~L., {Kumari}, N., {Emerick}, A., {et~al.} 2020, \mnras, 495, 2564

\bibitem[{{Kaasinen} {et~al.}(2018){Kaasinen}, {Kewley}, {Bian}, {Groves},
  {Kashino}, {Silverman}, \& {Kartaltepe}}]{kaasinen2018}
{Kaasinen}, M., {Kewley}, L., {Bian}, F., {et~al.} 2018, \mnras, 477, 5568

\bibitem[{{Kauffmann} {et~al.}(2003){Kauffmann}, {Heckman}, {White}, {Charlot},
  {Tremonti}, {Brinchmann}, {Bruzual}, {Peng}, {Seibert}, {Bernardi},
  {Blanton}, {Brinkmann}, {Castander}, {Cs{\'a}bai}, {Fukugita}, {Ivezic},
  {Munn}, {Nichol}, {Padmanabhan}, {Thakar}, {Weinberg}, \&
  {York}}]{Kauffmann2003}
{Kauffmann}, G., {Heckman}, T.~M., {White}, S. D.~M., {et~al.} 2003, \mnras,
  341, 33

\bibitem[{{Kauffmann} {et~al.}(2013){Kauffmann}, {Pillai}, \&
  {Goldsmith}}]{Kauffmann2013}
{Kauffmann}, J., {Pillai}, T., \& {Goldsmith}, P.~F. 2013, \apj, 779, 185

\bibitem[{{Kehrig} {et~al.}(2016){Kehrig}, {V{\'\i}lchez}, {P{\'e}rez-Montero},
  {Iglesias-P{\'a}ramo}, {Hern{\'a}ndez-Fern{\'a}ndez}, {Duarte Puertas},
  {Brinchmann}, {Durret}, \& {Kunth}}]{kehrig2016a}
{Kehrig}, C., {V{\'\i}lchez}, J.~M., {P{\'e}rez-Montero}, E., {et~al.} 2016,
  \mnras, 459, 2992

\bibitem[{{Kennicutt}(1998)}]{Kennicutt1998}
{Kennicutt}, Robert~C., J. 1998, \apj, 498, 541

\bibitem[{{Kennicutt} \& {Evans}(2012)}]{Kennicutt2012}
{Kennicutt}, R.~C. \& {Evans}, N.~J. 2012, \araa, 50, 531

\bibitem[{{Kewley} \& {Dopita}(2002)}]{kewley2002}
{Kewley}, L.~J. \& {Dopita}, M.~A. 2002, \apjs, 142, 35

\bibitem[{{Kewley} {et~al.}(2001){Kewley}, {Dopita}, {Sutherland}, {Heisler},
  \& {Trevena}}]{Kewley2001}
{Kewley}, L.~J., {Dopita}, M.~A., {Sutherland}, R.~S., {Heisler}, C.~A., \&
  {Trevena}, J. 2001, \apj, 556, 121

\bibitem[{Kewley \& Ellison(2008)}]{kewley2008}
Kewley, L.~J. \& Ellison, S.~L. 2008, 681, 1183

\bibitem[{{Kewley} {et~al.}(2006){Kewley}, {Groves}, {Kauffmann}, \&
  {Heckman}}]{kewley2006}
{Kewley}, L.~J., {Groves}, B., {Kauffmann}, G., \& {Heckman}, T. 2006, \mnras,
  372, 961

\bibitem[{Kewley {et~al.}(2019)Kewley, Nicholls, \& Sutherland}]{kewley2019}
Kewley, L.~J., Nicholls, D.~C., \& Sutherland, R.~S. 2019, 57, 511

\bibitem[{Kobulnicky \& Kewley(2004)}]{kobulnicky2004}
Kobulnicky, H.~A. \& Kewley, L.~J. 2004, 617, 240

\bibitem[{{Krumholz} \& {McKee}(2005)}]{Krumholz2005}
{Krumholz}, M.~R. \& {McKee}, C.~F. 2005, \apj, 630, 250

\bibitem[{{Krumholz} {et~al.}(2019){Krumholz}, {McKee}, \&
  {Bland-Hawthorn}}]{Krumholz2018}
{Krumholz}, M.~R., {McKee}, C.~F., \& {Bland-Hawthorn}, J. 2019, \araa, 57, 227

\bibitem[{{Kuncarayakti} {et~al.}(2018){Kuncarayakti}, {Anderson}, {Galbany},
  {Maeda}, {Hamuy}, {Aldering}, {Arimoto}, {Doi}, {Morokuma}, \&
  {Usuda}}]{kuncarayakti2018a}
{Kuncarayakti}, H., {Anderson}, J.~P., {Galbany}, L., {et~al.} 2018, \aap, 613,
  A35

\bibitem[{{Kunth} \& {{\"O}stlin}(2000)}]{kunth2000}
{Kunth}, D. \& {{\"O}stlin}, G. 2000, \aapr, 10, 1

\bibitem[{{Lagos} {et~al.}(2009){Lagos}, {Telles}, {Mu{\~n}oz-Tu{\~n}{\'o}n},
  {Carrasco}, {Cuisinier}, \& {Tenorio-Tagle}}]{lagos2009}
{Lagos}, P., {Telles}, E., {Mu{\~n}oz-Tu{\~n}{\'o}n}, C., {et~al.} 2009, \aj,
  137, 5068

\bibitem[{{Li} {et~al.}(2020){Li}, {Wang}, {Wu}, {Ma}, \& {Lin}}]{Li2020c}
{Li}, C., {Wang}, H.-C., {Wu}, Y.-W., {Ma}, Y.-H., \& {Lin}, L.-H. 2020,
  Research in Astronomy and Astrophysics, 20, 031

\bibitem[{{Lilly} {et~al.}(2013){Lilly}, {Carollo}, {Pipino}, {Renzini}, \&
  {Peng}}]{Lilly2013}
{Lilly}, S.~J., {Carollo}, C.~M., {Pipino}, A., {Renzini}, A., \& {Peng}, Y.
  2013, \apj, 772, 119

\bibitem[{{Liu} {et~al.}(2018){Liu}, {Wang}, {Lin}, {Gao}, {Liu}, {Berhane
  Teklu}, \& {Kong}}]{liu2018}
{Liu}, Q., {Wang}, E., {Lin}, Z., {et~al.} 2018, \apj, 857, 17

\bibitem[{{Lu} {et~al.}(2014){Lu}, {Zhao}, {Xu}, {Gao}, {Armus}, {Mazzarella},
  {Isaak}, {Petric}, {Charmandaris}, {D{\'\i}az-Santos}, {Evans}, {Howell},
  {Appleton}, {Inami}, {Iwasawa}, {Leech}, {Lord}, {Sanders}, {Schulz},
  {Surace}, \& {van der Werf}}]{lu2014}
{Lu}, N., {Zhao}, Y., {Xu}, C.~K., {et~al.} 2014, \apjl, 787, L23

\bibitem[{{Mainzer} {et~al.}(2014){Mainzer}, {Bauer}, {Cutri}, {Grav},
  {Masiero}, {Beck}, {Clarkson}, {Conrow}, {Dailey}, {Eisenhardt}, {Fabinsky},
  {Fajardo-Acosta}, {Fowler}, {Gelino}, {Grillmair}, {Heinrichsen}, {Kendall},
  {Kirkpatrick}, {Liu}, {Masci}, {McCallon}, {Nugent}, {Papin}, {Rice},
  {Royer}, {Ryan}, {Sevilla}, {Sonnett}, {Stevenson}, {Thompson}, {Wheelock},
  {Wiemer}, {Wittman}, {Wright}, \& {Yan}}]{mainzer2014}
{Mainzer}, A., {Bauer}, J., {Cutri}, R.~M., {et~al.} 2014, \apj, 792, 30

\bibitem[{{Mannucci} {et~al.}(2010){Mannucci}, {Cresci}, {Maiolino}, {Marconi},
  \& {Gnerucci}}]{Mannucci2010a}
{Mannucci}, F., {Cresci}, G., {Maiolino}, R., {Marconi}, A., \& {Gnerucci}, A.
  2010, \mnras, 408, 2115

\bibitem[{{Marino} {et~al.}(2013){Marino}, {Rosales-Ortega}, {S{\'a}nchez},
  {Gil de Paz}, {V{\'\i}lchez}, {Miralles-Caballero}, {Kehrig},
  {P{\'e}rez-Montero}, {Stanishev}, {Iglesias-P{\'a}ramo}, {D{\'\i}az},
  {Castillo-Morales}, {Kennicutt}, {L{\'o}pez-S{\'a}nchez}, {Galbany},
  {Garc{\'\i}a-Benito}, {Mast}, {Mendez-Abreu}, {Monreal-Ibero}, {Husemann},
  {Walcher}, {Garc{\'\i}a-Lorenzo}, {Masegosa}, {Del Olmo Orozco},
  {Mour{\~a}o}, {Ziegler}, {Moll{\'a}}, {Papaderos},
  {S{\'a}nchez-Bl{\'a}zquez}, {Gonz{\'a}lez Delgado}, {Falc{\'o}n-Barroso},
  {Roth}, {van de Ven}, \& {Califa Team}}]{Marino2013}
{Marino}, R.~A., {Rosales-Ortega}, F.~F., {S{\'a}nchez}, S.~F., {et~al.} 2013,
  \aap, 559, A114

\bibitem[{{Morisset} {et~al.}(2016){Morisset}, {Delgado-Inglada},
  {S{\'a}nchez}, {Galbany}, {Garc{\'\i}a-Benito}, {Husemann}, {Marino}, {Mast},
  \& {Roth}}]{Morisset2016}
{Morisset}, C., {Delgado-Inglada}, G., {S{\'a}nchez}, S.~F., {et~al.} 2016,
  \aap, 594, A37

\bibitem[{{Orlitova}(2020)}]{Orlitova2020}
{Orlitova}, I. 2020, arXiv e-prints, arXiv:2012.12378

\bibitem[{{{\"O}stlin} {et~al.}(2014){{\"O}stlin}, {Hayes}, {Duval},
  {Sandberg}, {Rivera-Thorsen}, {Marquart}, {Orlitov{\'a}}, {Adamo},
  {Melinder}, {Guaita}, {Atek}, {Cannon}, {Gruyters}, {Herenz}, {Kunth},
  {Laursen}, {Mas-Hesse}, {Micheva}, {Ot{\'\i}-Floranes}, {Pardy}, {Roth},
  {Schaerer}, \& {Verhamme}}]{ostlin2014}
{{\"O}stlin}, G., {Hayes}, M., {Duval}, F., {et~al.} 2014, \apj, 797, 11

\bibitem[{{Pan} {et~al.}(2019){Pan}, {Lin}, {Hsieh}, {Barrera-Ballesteros},
  {S{\'a}nchez}, {Hsu}, {Keenan}, {Tissera}, {Boquien}, {Dai}, {Knapen},
  {Riffel}, {Argudo-Fern{\'a}ndez}, {Xiao}, \& {Yuan}}]{pan2019}
{Pan}, H.-A., {Lin}, L., {Hsieh}, B.-C., {et~al.} 2019, \apj, 881, 119

\bibitem[{{Papadopoulos} {et~al.}(2007){Papadopoulos}, {Isaak}, \& {van der
  Werf}}]{Papadopoulos2007}
{Papadopoulos}, P.~P., {Isaak}, K.~G., \& {van der Werf}, P.~P. 2007, \apj,
  668, 815

\bibitem[{{Paudel} {et~al.}(2018){Paudel}, {Smith}, {Yoon},
  {Calder{\'o}n-Castillo}, \& {Duc}}]{Paudel2018}
{Paudel}, S., {Smith}, R., {Yoon}, S.~J., {Calder{\'o}n-Castillo}, P., \&
  {Duc}, P.-A. 2018, \apjs, 237, 36

\bibitem[{{Peng} {et~al.}(2010){Peng}, {Lilly}, {Kova{\v{c}}}, {Bolzonella},
  {Pozzetti}, {Renzini}, {Zamorani}, {Ilbert}, {Knobel}, {Iovino}, {Maier},
  {Cucciati}, {Tasca}, {Carollo}, {Silverman}, {Kampczyk}, {de Ravel},
  {Sanders}, {Scoville}, {Contini}, {Mainieri}, {Scodeggio}, {Kneib}, {Le
  F{\`e}vre}, {Bardelli}, {Bongiorno}, {Caputi}, {Coppa}, {de la Torre},
  {Franzetti}, {Garilli}, {Lamareille}, {Le Borgne}, {Le Brun}, {Mignoli},
  {Perez Montero}, {Pello}, {Ricciardelli}, {Tanaka}, {Tresse}, {Vergani},
  {Welikala}, {Zucca}, {Oesch}, {Abbas}, {Barnes}, {Bordoloi}, {Bottini},
  {Cappi}, {Cassata}, {Cimatti}, {Fumana}, {Hasinger}, {Koekemoer},
  {Leauthaud}, {Maccagni}, {Marinoni}, {McCracken}, {Memeo}, {Meneux}, {Nair},
  {Porciani}, {Presotto}, \& {Scaramella}}]{Peng2010}
{Peng}, Y.-j., {Lilly}, S.~J., {Kova{\v{c}}}, K., {et~al.} 2010, \apj, 721, 193

\bibitem[{Pettini \& Pagel(2004)}]{pettini2004}
Pettini, M. \& Pagel, B. E.~J. 2004, 348, L59

\bibitem[{{S{\'a}nchez} {et~al.}(2018){S{\'a}nchez}, {Avila-Reese},
  {Hernandez-Toledo}, {Cortes-Su{\'a}rez}, {Rodr{\'\i}guez-Puebla},
  {Ibarra-Medel}, {Cano-D{\'\i}az}, {Barrera-Ballesteros}, {Negrete},
  {Calette}, {de Lorenzo-C{\'a}ceres}, {Ortega-Minakata}, {Aquino},
  {Valenzuela}, {Clemente}, {Storchi-Bergmann}, {Riffel}, {Schimoia}, {Riffel},
  {Rembold}, {Brownstein}, {Pan}, {Yates}, {Mallmann}, \&
  {Bitsakis}}]{sanchez2018b}
{S{\'a}nchez}, S.~F., {Avila-Reese}, V., {Hernandez-Toledo}, H., {et~al.} 2018,
  \rmxaa, 54, 217

\bibitem[{S{\'a}nchez {et~al.}(2016{\natexlab{a}})S{\'a}nchez, P{\'e}rez,
  {S{\'a}nchez-Bl{\'a}zquez}, {Garc{\'i}a-Benito}, {Ibarra-Mede}, Gonz{\'a}lez,
  {Rosales-Ortega}, {S{\'a}nchez-Menguiano}, Ascasibar, Bitsakis, Law,
  {Cano-D{\'i}az}, {L{\'o}pez-Cob{\'a}}, Marino, {Gil de Paz},
  {L{\'o}pez-S{\'a}nchez}, {Barrera-Ballesteros}, Galbany, Mast,
  {Abril-Melgarejo}, \& {Roman-Lopes}}]{sanchez2016}
S{\'a}nchez, S.~F., P{\'e}rez, E., {S{\'a}nchez-Bl{\'a}zquez}, P., {et~al.}
  2016{\natexlab{a}}, Revista Mexicana de Astronomia y Astrofisica, 52, 171

\bibitem[{S{\'a}nchez {et~al.}(2016{\natexlab{b}})S{\'a}nchez, P{\'e}rez,
  {S{\'a}nchez-Bl{\'a}zquez}, Gonz{\'a}lez, {Ros{\'a}les-Ortega},
  {Cano-D{\'i}az}, {L{\'o}pez-Cob{\'a}}, Marino, {Gil de Paz}, Moll{\'a},
  {L{\'o}pez-S{\'a}nchez}, Ascasibar, \& {Barrera-Ballesteros}}]{sanchez2016a}
S{\'a}nchez, S.~F., P{\'e}rez, E., {S{\'a}nchez-Bl{\'a}zquez}, P., {et~al.}
  2016{\natexlab{b}}, Revista Mexicana de Astronomia y Astrofisica, 52, 21

\bibitem[{{Shangguan} {et~al.}(2019){Shangguan}, {Ho}, {Li}, {Zhuang}, {Xie},
  \& {Li}}]{Shangguan2019a}
{Shangguan}, J., {Ho}, L.~C., {Li}, R., {et~al.} 2019, \apj, 870, 104

\bibitem[{{Speagle} {et~al.}(2014){Speagle}, {Steinhardt}, {Capak}, \&
  {Silverman}}]{Speagle2014}
{Speagle}, J.~S., {Steinhardt}, C.~L., {Capak}, P.~L., \& {Silverman}, J.~D.
  2014, \apjs, 214, 15

\bibitem[{{Spence} {et~al.}(2018){Spence}, {Tadhunter}, {Rose}, \&
  {Rodr{\'\i}guez Zaur{\'\i}n}}]{Spence2018}
{Spence}, R.~A.~W., {Tadhunter}, C.~N., {Rose}, M., \& {Rodr{\'\i}guez
  Zaur{\'\i}n}, J. 2018, \mnras, 478, 2438

\bibitem[{{Steffen} {et~al.}(2021){Steffen}, {Fu}, {Comerford}, {Dai}, {Feng},
  {Gross}, \& {Xue}}]{steffen2021}
{Steffen}, J.~L., {Fu}, H., {Comerford}, J.~M., {et~al.} 2021, \apj, 909, 120

\bibitem[{{Stierwalt} {et~al.}(2015){Stierwalt}, {Besla}, {Patton}, {Johnson},
  {Kallivayalil}, {Putman}, {Privon}, \& {Ross}}]{Stierwalt2015}
{Stierwalt}, S., {Besla}, G., {Patton}, D., {et~al.} 2015, \apj, 805, 2

\bibitem[{{Teklu} {et~al.}(2020){Teklu}, {Gao}, {Kong}, {Lin}, \&
  {Liang}}]{Teklu2020a}
{Teklu}, B.~B., {Gao}, Y., {Kong}, X., {Lin}, Z., \& {Liang}, Z. 2020, \apj,
  897, 61

\bibitem[{{Tremonti} {et~al.}(2004){Tremonti}, {Heckman}, {Kauffmann},
  {Brinchmann}, {Charlot}, {White}, {Seibert}, {Peng}, {Schlegel}, {Uomoto},
  {Fukugita}, \& {Brinkmann}}]{Tremonti2004}
{Tremonti}, C.~A., {Heckman}, T.~M., {Kauffmann}, G., {et~al.} 2004, \apj, 613,
  898

\bibitem[{{Yao} {et~al.}(2022){Yao}, {Chen}, {Liu}, {Chen}, {Lin}, {Zhang},
  {Gao}, \& {Kong}}]{yao2022a}
{Yao}, Y., {Chen}, G., {Liu}, H., {et~al.} 2022, \aap, 661, A112

\bibitem[{{Zaragoza-Cardiel} {et~al.}(2014){Zaragoza-Cardiel}, {Font},
  {Beckman}, {Garc{\'\i}a-Lorenzo}, {Erroz-Ferrer}, \&
  {Guti{\'e}rrez}}]{zaragoza-cardiel2014}
{Zaragoza-Cardiel}, J., {Font}, J., {Beckman}, J.~E., {et~al.} 2014, \mnras,
  445, 1412

\bibitem[{{Zhang} {et~al.}(2020{\natexlab{a}}){Zhang}, {Paudel}, {Smith},
  {Duc}, {Puzia}, {Peng}, {C{\^o}te}, {Ferrarese}, {Boselli}, {Wang}, \&
  {Oh}}]{Zhang2020h}
{Zhang}, H.-X., {Paudel}, S., {Smith}, R., {et~al.} 2020{\natexlab{a}}, \apjl,
  891, L23

\bibitem[{{Zhang} {et~al.}(2020{\natexlab{b}}){Zhang}, {Smith}, {Oh}, {Paudel},
  {Duc}, {Boselli}, {C{\^o}t{\'e}}, {Ferrarese}, {Gao}, {Hunter}, {Puzia},
  {Peng}, {Rong}, {Shin}, \& {Zhao}}]{Zhang2020g}
{Zhang}, H.-X., {Smith}, R., {Oh}, S.-H., {et~al.} 2020{\natexlab{b}}, \apj,
  900, 152

\bibitem[{{Zhang} {et~al.}(2014){Zhang}, {Gao}, {Henkel}, {Zhao}, {Wang},
  {Menten}, \& {G{\"u}sten}}]{zhang2014}
{Zhang}, Z.-Y., {Gao}, Y., {Henkel}, C., {et~al.} 2014, \apjl, 784, L31

\bibitem[{{Zou} {et~al.}(2019){Zou}, {Gao}, {Zhou}, \& {Kong}}]{zou2019}
{Zou}, H., {Gao}, J., {Zhou}, X., \& {Kong}, X. 2019, \apjs, 242, 8

\end{thebibliography}

\end{document}